\documentclass[sigplan,screen]{acmart}

\usepackage[utf8]{inputenc} 
\usepackage{tabularx}
\usepackage[T1]{fontenc}    
\usepackage{url}            
\usepackage{booktabs}       
\usepackage{amsfonts}       
\usepackage{nicefrac}       
\usepackage{microtype}      
\usepackage{xcolor}         

\usepackage{float}
\usepackage{graphicx}
\usepackage{multirow}
\usepackage{subcaption}
\usepackage{amsmath}
\usepackage{xcolor}
\usepackage{rotating}
\usepackage{array}
\usepackage{makecell}
\usepackage{tabularx}
\usepackage{listings}
\newcommand{\ours}[0]{\textsc{DreamCraft}}
\newcommand{\unconstrained}[0]{\textsc{Unconstrained NeRF}}

\newcommand{\ie}{\textit{i.e.,}}

\def\*#1{\mathbf{#1}}

\AtBeginDocument{%
  }

\copyrightyear{2024}
\acmYear{2024}
\setcopyright{acmlicensed}\acmConference[FDG 2024]{Proceedings of the 19th International Conference on the Foundations of Digital Games}{May 21--24, 2024}{Worcester, MA, USA}
\acmBooktitle{Proceedings of the 19th International Conference on the Foundations of Digital Games (FDG 2024), May 21--24, 2024, Worcester, MA, USA}
\acmDOI{10.1145/3649921.3649943}
\acmISBN{979-8-4007-0955-5/24/05}

\begin{document}

\title{\ours{}: Text-Guided Generation \\ of Functional 3D Environments in Minecraft}


\author{Sam Earle}
\affiliation{%
  \institution{New York University}
  \city{Broklyn}
  \country{USA}
}

\author{Filippos Kokkinos}
\affiliation{%
 \institution{Meta}
 \city{London}
 \country{UK}}
 
\author{Yuhe Nie}
\affiliation{%
 \institution{New York University}
 \city{Brooklyn}
 \country{USA}}
 
\author{Julian Togelius}
\affiliation{%
 \institution{New York University}
 \city{Brooklyn}
 \country{USA}}
 
\author{Roberta Raileanu}
\affiliation{%
 \institution{Meta}
 \city{London}
 \country{UK}}








\begin{abstract}
Procedural Content Generation (PCG) algorithms enable the automatic generation of complex and diverse artifacts. However, they don't provide high-level control over the generated content and typically require domain expertise. In contrast, text-to-3D methods allow users to specify desired characteristics in natural language, offering a high amount of flexibility and expressivity. But unlike PCG, such approaches cannot guarantee functionality, which is crucial for certain applications like game design. 
In this paper, we present a method for generating functional 3D artifacts from free-form text prompts in the open-world game Minecraft. Our method, \ours{}, trains quantized Neural Radiance Fields (NeRFs) to represent artifacts that, when viewed in-game, match given text descriptions.
We find that \ours{} produces more aligned in-game artifacts than a baseline that post-processes the output of an unconstrained NeRF. 
Thanks to the quantized representation of the environment, functional constraints can be integrated using specialized loss terms. We show how this can be leveraged to generate 3D structures that match a target distribution or obey certain adjacency rules over the block types. 
\ours{} inherits a high degree of expressivity and controllability from the NeRF, while still being able to incorporate functional constraints through domain-specific objectives.
\end{abstract}




\settopmatter{printfolios=true}

\maketitle

\begin{figure*}
\begin{subfigure}[b]{\textwidth}
\begin{subfigure}[b]{.32\textwidth}
\includegraphics[width=\textwidth]{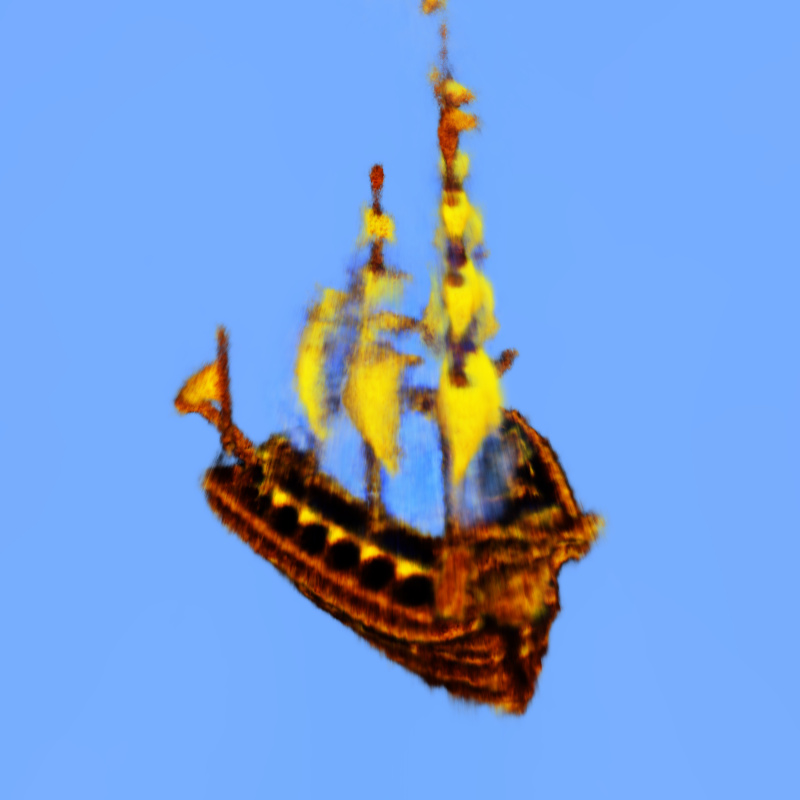}
\caption*{\unconstrained}
\end{subfigure}
\begin{subfigure}[b]{.32\textwidth}
\includegraphics[width=\textwidth]{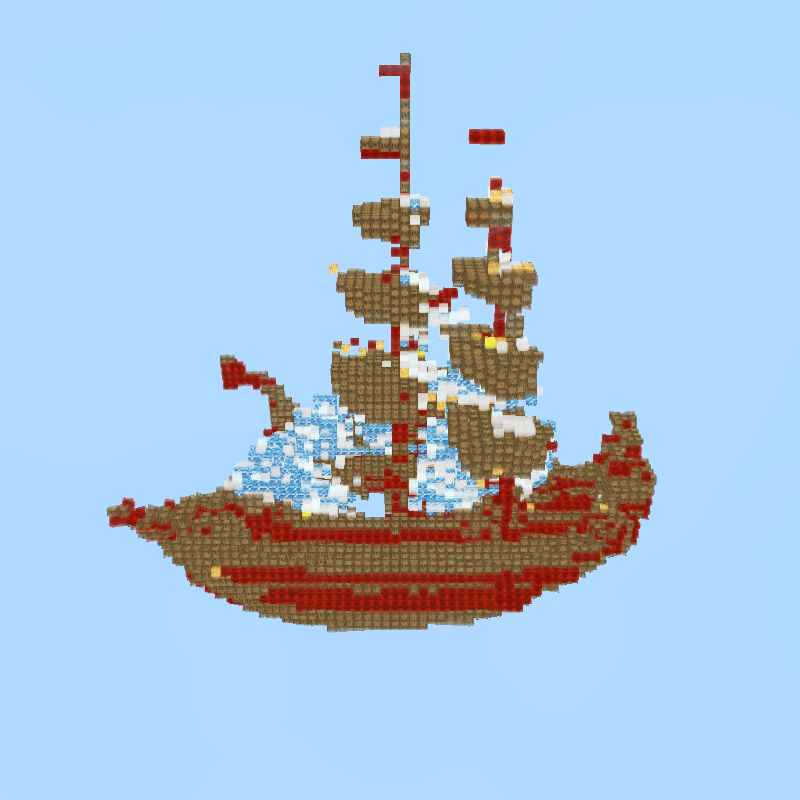}
\caption*{\ours{} Neural Render}
\end{subfigure}
\begin{subfigure}[b]{.32\textwidth}
\includegraphics[width=\textwidth]{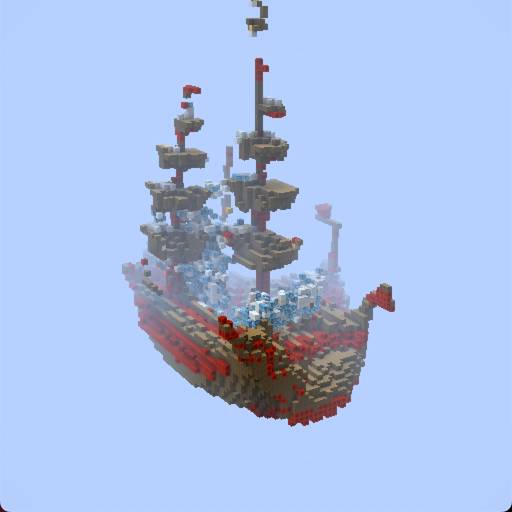}
\caption*{\ours{} In-Game}
\end{subfigure}
\caption{\textit{large medieval ship}}
\end{subfigure}\\
\begin{subfigure}[b]{\textwidth}
\begin{subfigure}[b]{.32\textwidth}
\includegraphics[width=\textwidth]{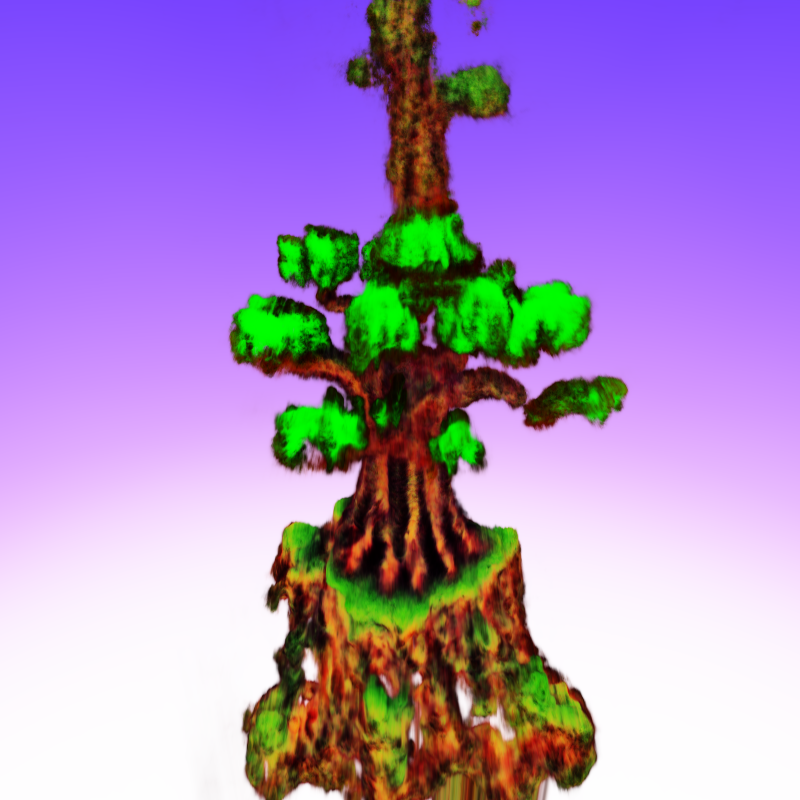}
\caption*{\unconstrained{}}
\end{subfigure}
\begin{subfigure}[b]{.32\textwidth}
\includegraphics[width=\textwidth]{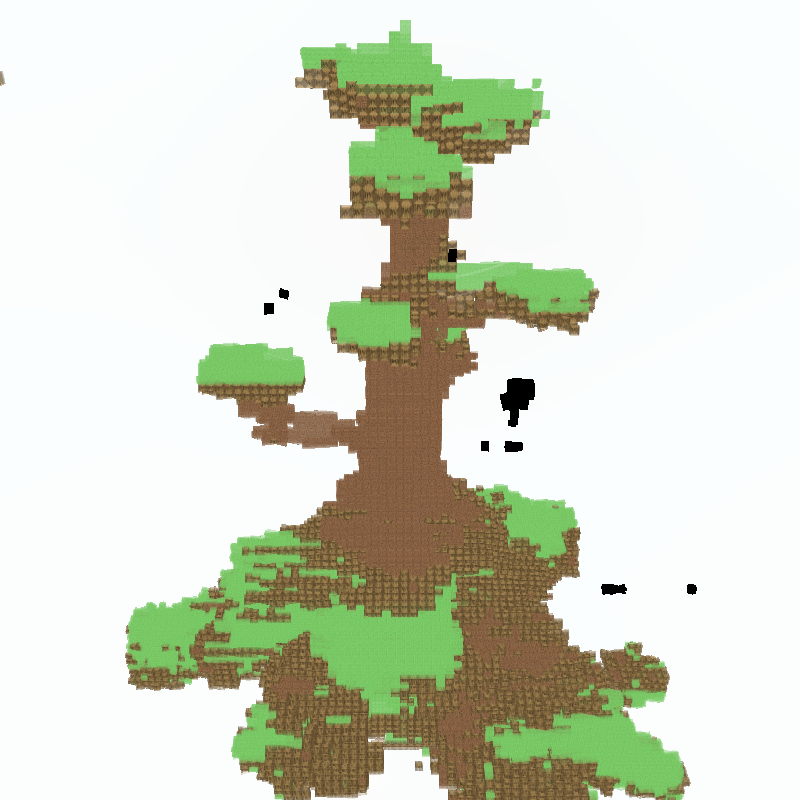}
\caption*{\ours{} Neural Render}
\end{subfigure}
\begin{subfigure}[b]{.32\textwidth}
\includegraphics[width=\textwidth]{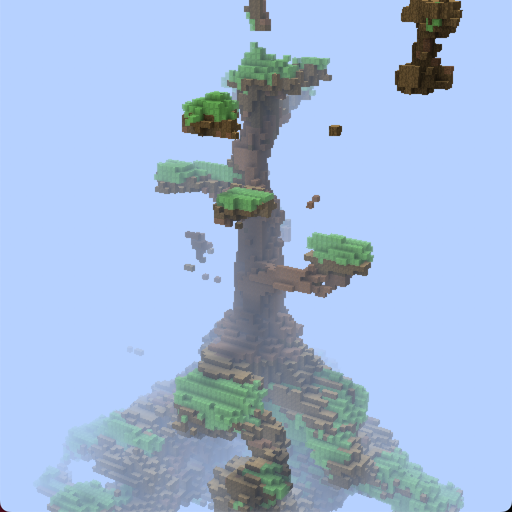}
\caption*{\ours{} In-Game}
\end{subfigure}
\caption{\textit{old wizards tree mansion series 2 build 1 read description}}
\end{subfigure}
\caption{Comparison of unconstrained NeRF (left), \ours{} neural render, and \ours{} in-game generation for a Planet Minecraft caption. Despite its lower resolution, \ours{}'s generated structures are similar in quality to those of the unconstrained NeRF, both closely matching the corresponding textual description.}
\label{fig:planet_mc_unconstrained}
\end{figure*}

\section{Introduction}


Procedural Content Generation (PCG) refers to a class of algorithms that can automatically create content such as video game levels~\cite{shaker2016procedural,hendrikx2013procedural,risi2020increasing,khalifa2020pcgrl,dahlskog2014procedural,liu2021deep}, 2D or 3D visual assets~\cite{watkins2016procedural,liu2021deep,brocchini2022monster,nair2020using,liapis2013designer,sudhakaran2021growing}, game rules or mechanics~\cite{summerville2018procedural,nelson2016rules,togelius2011search,gravina2019procedural,zook2014generating}, or reinforcement learning (RL) environments~\cite{justesen2018illuminating,juliani2019obstacle,risi2020increasing,kuttler2020nethack,dennis2020emergent,samvelyan2021minihack,team2021open,jiang2021prioritized,gisslen2021adversarial,bontrager2021learning,team2023human,jiang2022learning,parker2022evolving}. PCG allows for compression of information~\cite{summerville2015sampling,togelius2011search}, increased replayability via endless variation~\cite{yannakakis2011experience,smith2011answer,brewer2017computerized}, expression of particular aesthetics~\cite{liapis2012adapting,alvarez2018assessing,canossa2015towards,guzdial2017visual}, and reduction of human labour otherwise required to manually produce artifacts~\cite{shaker2010towards,gao2022procedural,dieterich2017using,shaker2016mixed}. 
These methods are procedural in the sense that they outline sets of procedures or rules for generating artifacts, such as adjacency constraints in Wave Function Collapse, local update rules in cellular automata, or heuristic search in constraint satisfaction.
These procedures often leverage domain-specific knowledge in order to guarantee that generated artifacts are functional; that a game environment does not contain structures that violate physics, or that a player is able to navigate between key points within them.
However, users cannot generally control such methods via free-form language, and control is limited to those metrics explicitly defined by designers.


In contrast, recent generative models have shown impressive abilities in generating diverse images, videos, or 3D scenes from text prompts describing the desired output in natural language~\cite{Rombach_2022_CVPR,poole2022dreamfusion,singer2022make}. These advances allow users to create high-quality content even if they are not domain experts. While these models can produce controllable and open-ended generations, the created content is not guaranteed to be functional. Functionality is particularly important for certain applications such as game design or the creation of RL environments. 
Some recent efforts leverage language models to generate level representations~\cite{todd2023level,sudhakaran2023mariogpt}, but effectively reduce text-based controls to a series of scalar values.
Other methods train generative text-image models to produce levels that are made of discrete tiles~\cite{merino2023five} or controllable by actions~\cite{bruce2024genie}, but they do not bring any functional guarantees: houses may spawn in disjointed, and birds may turn into bumblebees after a few frames.

\begin{figure*}[ht]
  \begin{subfigure}[t]{0.32\textwidth}
    \centering
    \includegraphics[width=\textwidth]{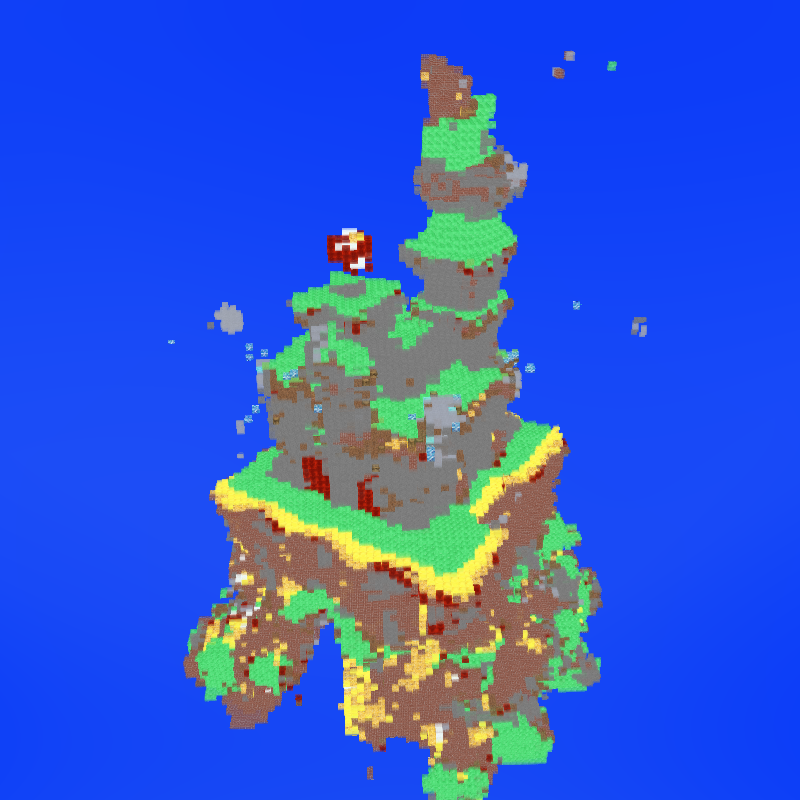}
    \caption{\textit{mario kartgba bowsers castle 2}}
  \end{subfigure}
  \hfill
  \begin{subfigure}[t]{0.32\textwidth}
    \centering
    \includegraphics[width=\textwidth,trim={120 120 120 120},clip]{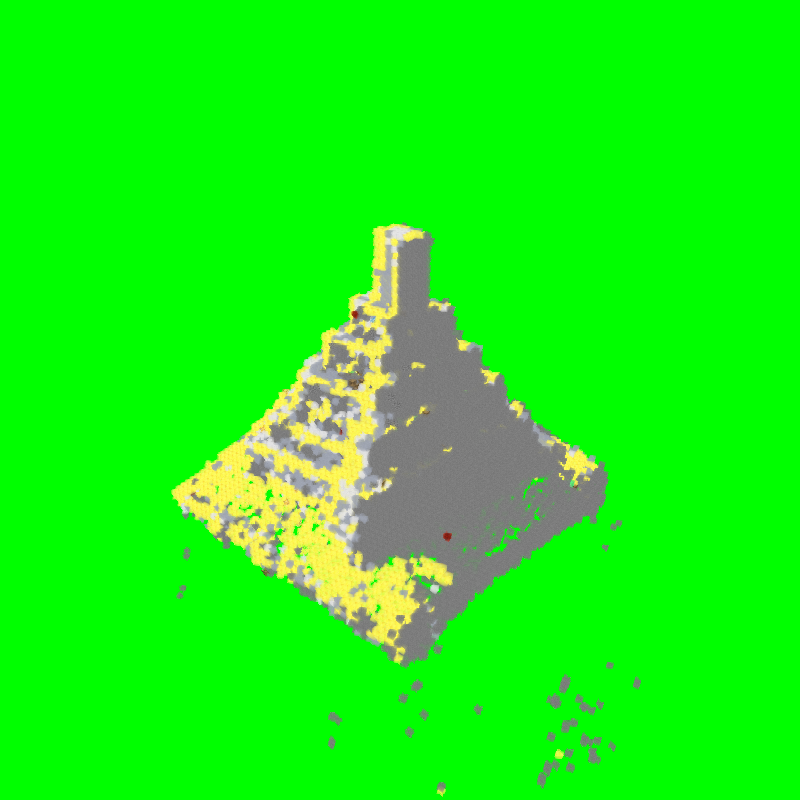}
    \caption{\textit{the ziggurat}}
  \end{subfigure}
  \hfill
  \begin{subfigure}[t]{0.32\textwidth}
    \centering
    \includegraphics[width=\textwidth]{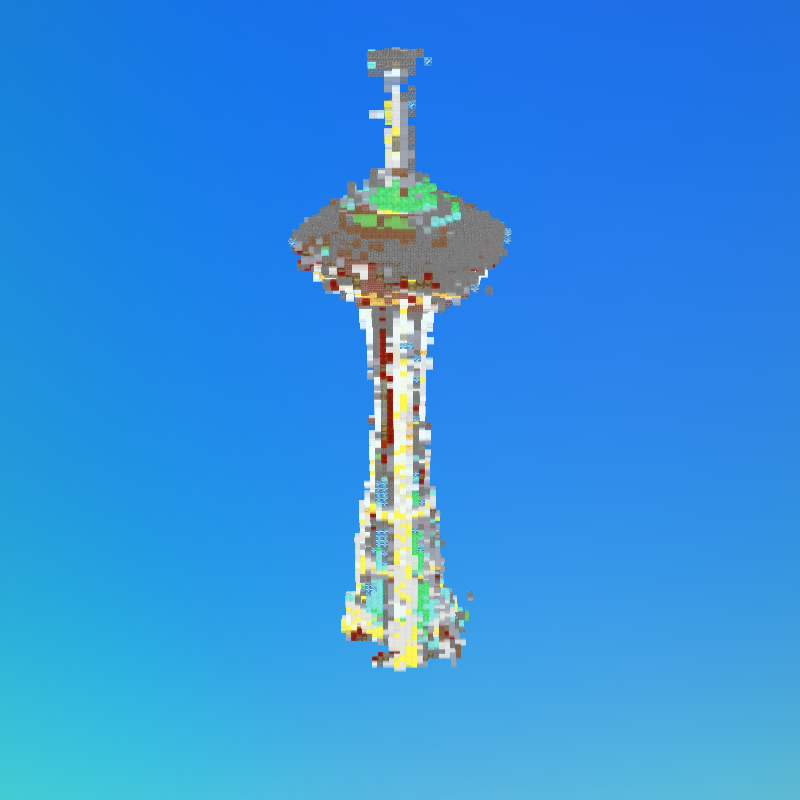}
    \caption{\textit{space needle}}
  \end{subfigure}
  \hfill
  \begin{subfigure}[t]{0.32\textwidth}
    \centering
    \includegraphics[width=\textwidth,trim={120 120 120 120},clip]{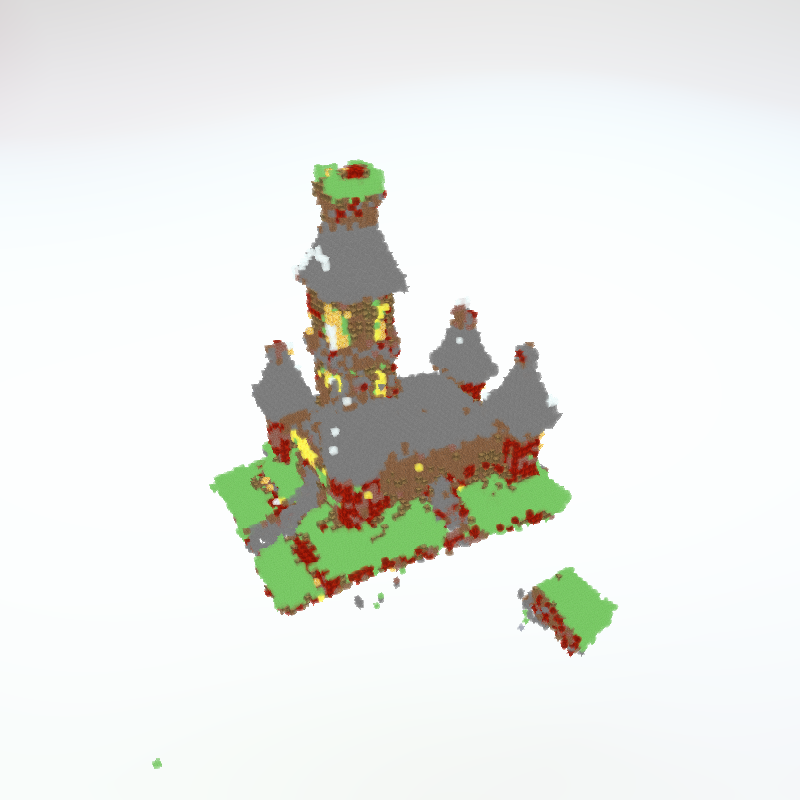}
    \caption{\textit{shubbles castle building contest}}
  \end{subfigure}
  \hfill 
  \begin{subfigure}[t]{0.32\textwidth}
    \centering
    \includegraphics[width=\textwidth]{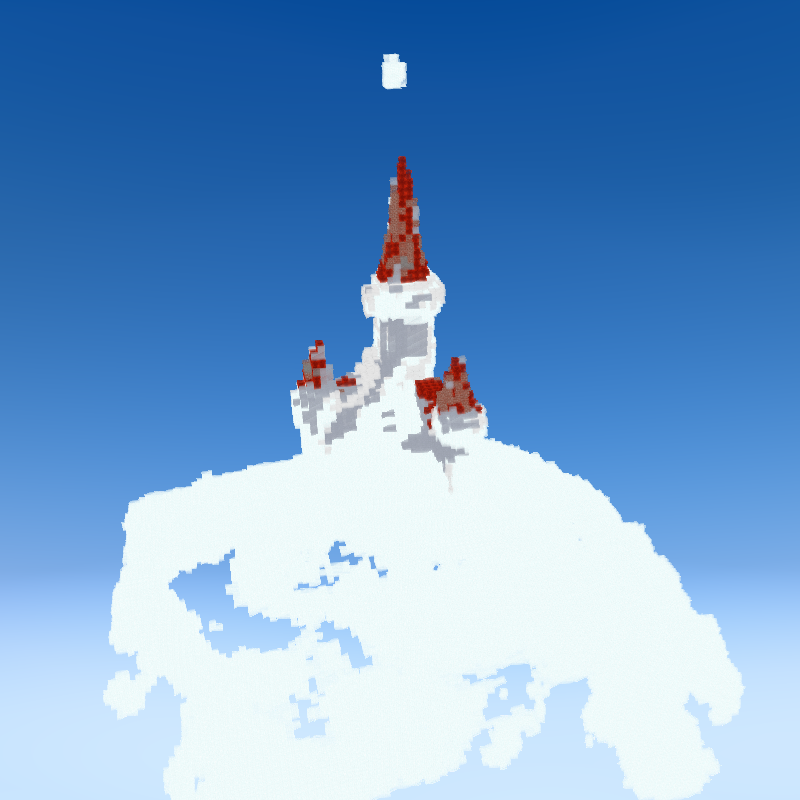}
    \caption{\textit{white snowy castle}}
  \end{subfigure}
  \hfill 
  \begin{subfigure}[t]{0.32\textwidth}
    \centering
    \includegraphics[width=\textwidth]{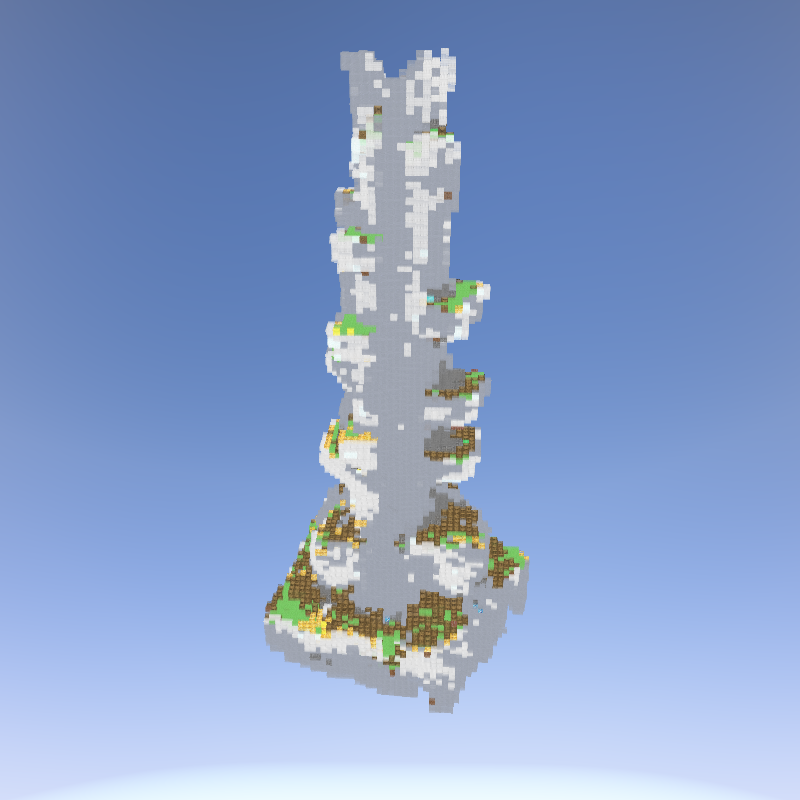}
    \caption{\textit{quartz tower 1}}
  \end{subfigure}
  \hfill
  \begin{subfigure}[t]{0.32\textwidth}
    \centering
    \includegraphics[width=\textwidth,trim={150 150 150 150},clip]{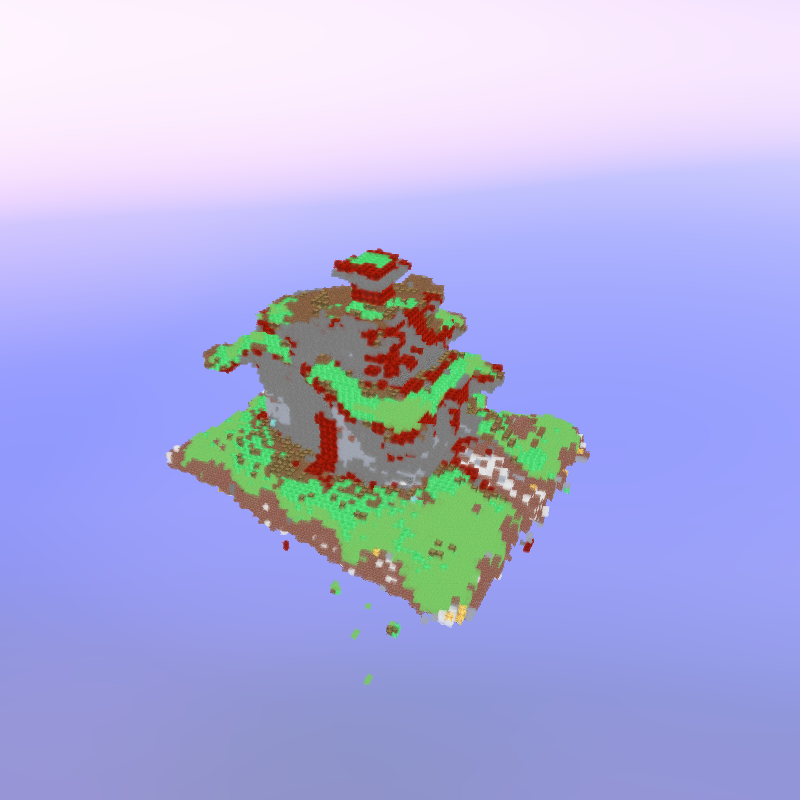}
    \caption{\textit{rustic fantasy house timelapse download}}
  \end{subfigure}
  \hfill
  \begin{subfigure}[t]{0.32\textwidth}
    \centering
    \includegraphics[width=\textwidth,trim={200 200 200 200},clip]{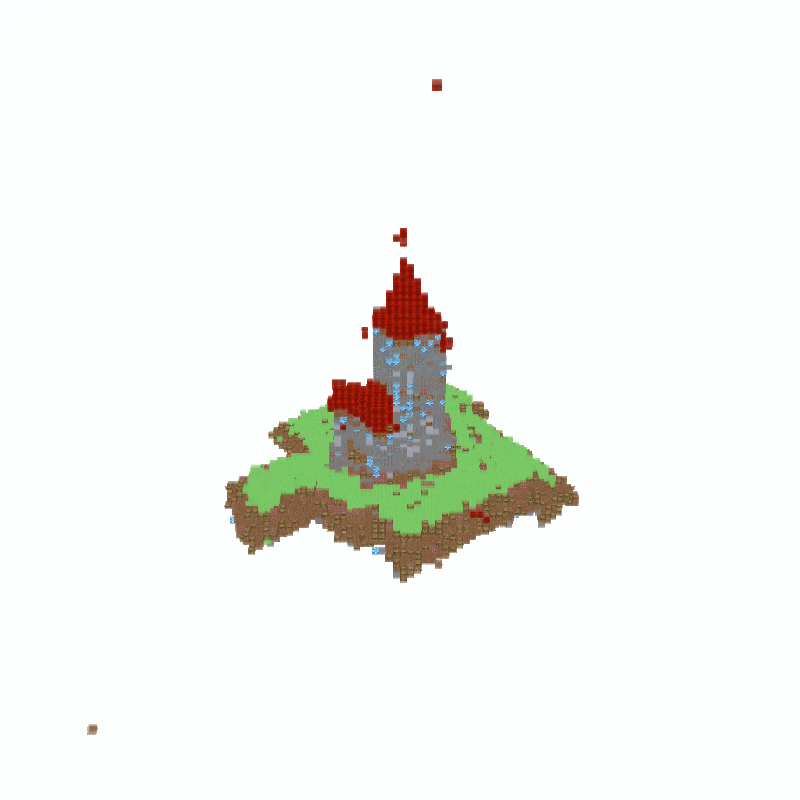}
    \caption{\textit{small castle}}
  \end{subfigure}
  \hfill
  \begin{subfigure}[t]{0.32\textwidth}
    \centering
    \includegraphics[width=\textwidth,trim={100 100 100 100},clip]{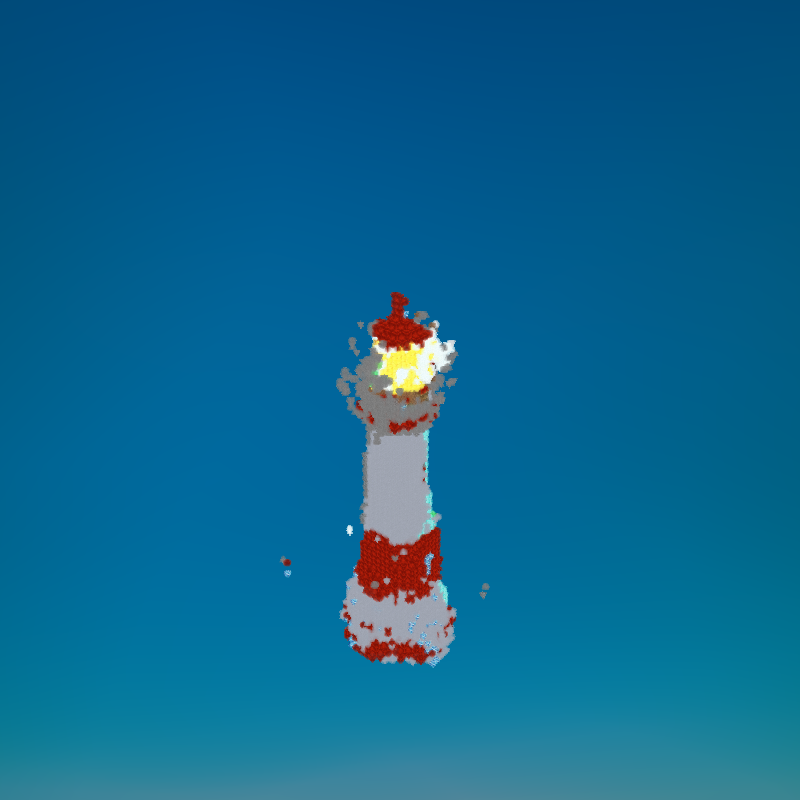}
    \caption{\textit{space lighthouse}}
  \end{subfigure}
  \caption{\ours{} neural rendered generations for a set of Planet Minecraft captions.}
\label{fig:planet_minecraft_generations}
\end{figure*}

In this work, we pursue a hybrid approach, adapting a generative model to operate on discrete 3D assets and incorporate functional constraints into its loss function.
We propose a new method for generating functional 3D environments from free-form text prompts in the open-world game Minecraft. Our method, \ours{}, trains a quantized NeRF to produce an environment layout that, when viewed in-game, matches a given text description (see Figures~\ref{fig:planet_mc_unconstrained} and~\ref{fig:planet_minecraft_generations} for some examples). 
Experimenting with various quantization schemes, we find that using soft air blocks or annealing them from continuous (soft) to discrete is crucial for learning stability, while using discrete block types leads to the most recognizable structures. We evaluate the fidelity of our approach in matching generated artifacts to descriptions of both generic and domain-specific scenes and objects. We find that \ours{} produces in-game artifacts that align with inputs more consistently than a baseline that post-processes the output of an unconstrained NeRF. 

Thanks to its quantized representation of the game world, \ours{} can jointly optimize loss terms that enforce local functional constraints on patterns of blocks. We show how this can be instantiated to, for example, generate 3D structures that match a target distribution or obey certain adjacency rules over the block types. 
By inheriting a high degree of expressivity and controllability from the NeRF, while still being able to incorporate functional constraints through domain-specific objectives, \ours{} combines the strengths of both PCG and generative AI approaches, representing a first step towards democratizing flexible yet functional content creation. Our method has potential applications in the development of AI assistants for game design, as well as in the production of diverse and controllable environments for training and evaluating RL agents.

To summarize, our paper makes the following contributions:
\begin{enumerate}
\item introduces \ours{}, a new method for training a quantized NeRF to produce 3D structures that match a given textual description using a set of discrete Minecraft blocks,
\item studies different quantization schemes such as whether to use discrete or continuous block densities and types, 
\item shows that the quantized NeRF produces more accurate Minecraft artifacts than an unconstrained NeRF, and
\item demonstrates how to incorporate functional constraints such as obeying certain target block distributions or adjacency rules.
\end{enumerate}

\section{Related Work}

\textbf{Procedural Content Generation (PCG)} is becoming increasingly more popular for training and evaluating robust RL agents that can generalize across a wide range of settings~\cite{justesen2018illuminating,juliani2019obstacle,kuttler2020nethack,dennis2020emergent,samvelyan2021minihack,jiang2021prioritized,team2021open,team2023human}. 
Generative models like ours provide a way of biasing the environment generations towards human-relevant ones, thus enabling to search more efficiently through vast environment spaces. Existing works indicate that environment generation can be controlled via computable metrics~\cite{earle2022illuminating, earle2021learning, khalifa2020pcgrl, jiang2022learning, green2020mario, sarkar2020sequential, sarkar2020conditional, mott2019controllable, shaker2010towards}.
Novel environments can also be generated by learning on human datasets~\cite{siper2022path, guzdial2022pcgml, guzdial2022procedural, liu2021deep, lopez2020deep, summerville2018procedural}, sometimes with additional functional constraints or post-processing~\cite{guzdial2022constraint, zhang2020video, lee2020precomputing, torrado2020bootstrapping, karth2019addressing}. More recently,~\citet{todd2023level,sudhakaran2023mariogpt} use large language models to generate Sokoban and Mario 2D levels. However, our work is first to show how multi-modal models can be leveraged to guide the generation of 3D game environments. One of the most popular PCG algorithms is wave function collapse~\cite{Gumin_Wave_Function_Collapse_2016} which generates structurally consistent content from a single sample, such that its output matches tile-frequency and adjacency constraints. In this paper, we show how such constraints can be used in conjunction with text-guidance to generate environments where both high-level (e.g. via natural language) and low-level (e.g. via block target distributions or adjacency rules) aspects can be controlled.

\textbf{Text-to-3D Generation} Our work builds upon the many recent advances in text-to-3D generation~\cite{poole2022dreamfusion,mildenhall2020nerf,chen2022tensorf}. For example, DVGO~\cite{sun2022direct} is a supervised NeRF method that, instead of training an multi-layer perceptron (MLP), directly optimizes a voxel grid over the 3D space. 
Similarly, PureCLIPNeRF~\cite{lee2022understanding} uses a CLIP loss to guide a NeRF using both direct and implicit voxel grids. Our approach, \ours{}, resembles an implicit voxel grid approach, in that it uses MLPs to parameterize activations over discrete grids. 
But instead of outputting continuous RGB and density values and interpolating between nearest grid vertices (to determine activation at a given point during ray tracing), our approach uses MLPs to produce predictions over block types, and considers only the single nearest grid vertex during ray sampling (to determine within which block a given point resides).

\textbf{Text-to-Environment Generation}
\cite{merino2023five} train a \\text-conditioned decoder model on a dataset of hand-made levels in a 2D tile domain.
Compared to NeRFs, which train a model to represent a single artifact (here, a level), the ``5 dollar" decoder model can represent a distribution of artifacts.
It can also potentially achieve a certain amount of generalization thanks to the pre-trained LLM which is used to encode text prompts.

\textbf{Minecraft Environment Generation} Several prior works have sought to generate Minecraft environments using both supervised and self-supervised methods. \citet{awiszus2021world} use a 3D GAN~\cite{goodfellow2020generative} architecture to generate arbitrarily sized world snippets from a single example. Meanwhile, \citet{hao2021gancraft} envision Minecraft as a potential ``sketchbook'' for designing more photorealistic 3D landscapes, using an unsupervised neural rendering approach to generate the latter from preexisting Minecraft landscapes.
To assist Minecraft players, \citet{merino2023interactive} introduce a tool for interactive evolution using both a 3D generative model to generate the structure design and an encoding model for applying Minecraft-specific textures to the structure's voxels. \citet{sudhakaran2021growing} have shown that neural cellular automata can be used to grow complex 3D Minecraft artifacts made out of thousands of blocks such as castles, apartment blocks, and trees. Other works have leveraged search-based methods~\cite{yates2021use}, evolutionary algorithms~\cite{medina2023evolving,skjeltorp20223d}, or even reinforcement learning~\cite{jiang2022learning} approaches to generate Minecraft structures. The game has also been a testbed for PCG algorithms~\cite{salge2018generative,salge2022impressions}, open-endedness~\cite{grbic2021evocraft}, artificial life~\cite{sudhakaran2021growing}, RL agents~\cite{johnson2016malmo, milani2020retrospective, guss2021minerl, kanervisto2022minerl}, or foundation models for decision-making~\cite{pmlr-v176-kanervisto22a,fan2022minedojo,baker2022video,wang2022attention}. However, our work is first to generate functional Minecraft environments directly from text prompts, enhancing the high-level controllability of the creation process.





\section{\ours{}: Text-Guided Minecraft Environment Generation}

\begin{figure*}
\includegraphics[width=\linewidth]{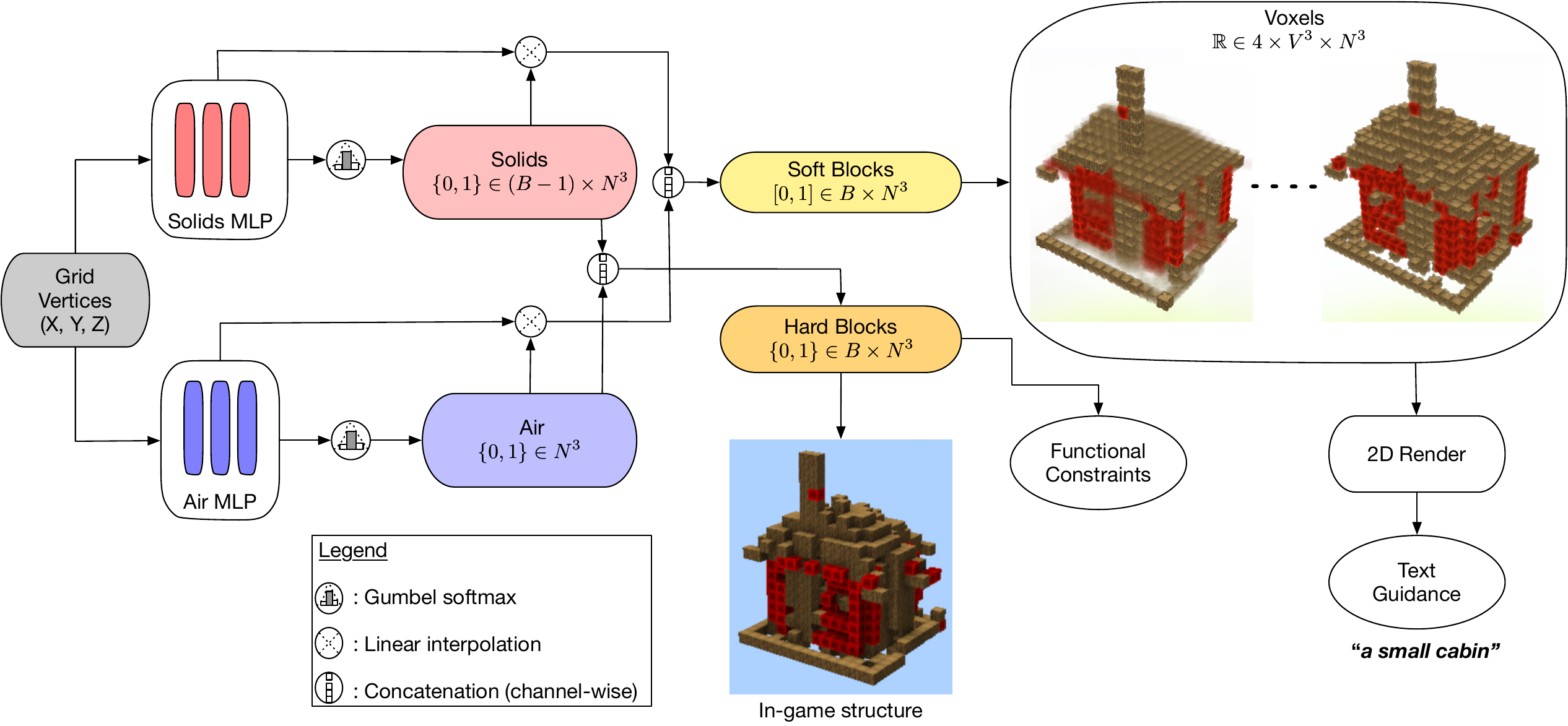}
\caption{\ours{} produces a distribution of discrete blocks over a grid. Unquantized or ``soft" versions of this distribution can be visualized for text-image guidance, while a fully discrete representation can be used to determine functional constraints, and exported to an in-game structure.}
\label{fig:overview}
\end{figure*}






\subsection{Quantized NeRFs for Environment Generation}

In this section, we introduce \ours{}, a quantized NeRF which learns to arrange in-game assets during training (see Figure~\ref{fig:overview} for an overview of our approach).
Our text-guided NeRF implementation uses score distillation sampling from a pre-trained image generation model to provide a loss function for the optimization of the NeRF.
We use a preliminary version of Emu~\cite{dai2023emu}, trained only on Shutterstock data.
Note that our approach is agnostic to the text-to-3D model used, so it can be applied to any other text-to-3D model architecture and thus benefit from future advances in that field. Also note that we do not use any additional or domain-specific data to train our model apart from the block textures in Table~\ref{tab:textures}. 

\begin{figure*}[t]
\includegraphics[width=\linewidth]{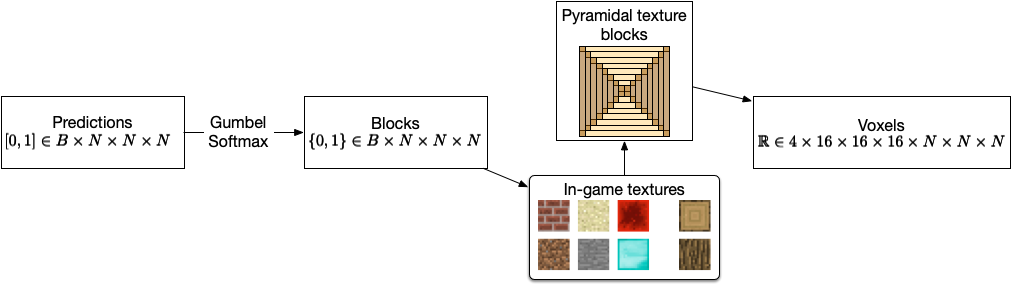}
\caption{\textbf{From 2D textures to 3D structures} Once discretized, block types are mapped to voxel grids corresponding to the appearance of objects in-game. Here, we assemble each block into a 3D voxel grid using 2D textures from Minecraft. We approximate solid blocks by repeating the surface texture in the space inside the block.}
\label{fig:pyramidal_textures}
\end{figure*}

\textbf{From Continuous Points to Discrete Blocks} As in PureCLIPNeRF, we use MLPs to predict continuous activations over a grid by feeding them $X, Y, Z$ coordinates which are encoded using a series of sine waves, as in the original NeRF~\cite{mildenhall2020nerf}. 
We sample $X, Y, Z$ vertices along a cubic $3D$ grid of width $N$, generating predictions over solid block types at each cell in the grid.
We refer to the resulting tensor as the soft \textit{solid block grid} $\*B_{\mathrm{soft}} \in \left[0, 1 \right]^{M \times N\times N\times N}$, of width $N$, comprising of $M$ different non-air block types.
We take gumbel softmax~\cite{jang2017categorical} over these predictions to yield a discrete grid of solid blocks $\*B_{\mathrm{hard}}$ with values in $[0, 1]$.
We feed the same $X, Y, Z$ coordinates to a separate MLP to generate a soft \textit{air block grid} $\*A_{\mathrm{soft}}\in \left[0,1\right]^{N\times N\times N}$, and again apply gumbel softmax to generate its discrete counterpart $\*A_{\mathrm{hard}} \in \left\{0, 1\right\}^{N\times N\times N}$. The distinction between air and solid blocks is analogous to that between the air and albedo MLPs in standard NeRFs.

During training, we may interpolate hard and soft variants to interpolate final air and block grids:
\begin{align*}
\*A= \alpha \cdot \*A_{\mathrm{hard}} + (1-\alpha)\cdot \*A_{\mathrm{soft}} && \*B = \beta \cdot \*B_{\mathrm{hard}} + (1 - \beta)\cdot \*B_{\mathrm{soft}},
\end{align*} where $\alpha$ and $\beta$ in $[0,1]$ control the hardness of air and solid blocks respectively. In our experiments, we set these to $1$ and $0$ when air/solids are hard (discrete) and soft (continuous) respectively, and linearly scale between $0$ and $1$ to ``anneal'' the blocks from continuous to discrete. We then mask solid block activations with air activations to yield the final block grid $\*C = \*A \odot \*B$, with $\*C \in [0,1]^{M\times N\times N\times N}$, where $\odot$ denotes the element-wise product, broadcasting over the block type dimension in $\*B$. Thus, $\*C$ can be seen as a 3D image with as many channels as there are solid block types. When a cell in $\*C$ has a value of $1$ in a given block channel, and $0$s elsewhere, it contains only this block type. When it has values in $(0,1)$, a block is ``partially'' present. And when a cell in $\*C$ is all zeros, it represents the presence of an air block. When exporting structures to the game engine or computing loss from functional constraints, we set $\alpha=\beta=1$, producing an entirely discrete block grid $\*C_{\mathrm{hard}}$.

\textbf{From 2D Textures to 3D Structures} 

The block grid $\*C$ is combined with statically-generated voxel grids to differentiably generate artifacts that visually resemble in-game structures.
First, we pre-fabricate $16\times 16\times 16$ voxel grids for each block type using in-game textures, applying the game's $16\times 16$ RGB textures to the appropriate block faces (picking an arbitrary priority order among faces where voxels overlap along the edges of the cube).
These grids are frozen and do not pass gradients during learning.
During the forward pass, we project our $M\times N \times N\times N$ block grid into a $16N\times 16N \times 16N$ voxelated block grid, where blocks appear in the same arrangement as in the low-resolution block grid, but in their voxelated form.
We then apply neural ray tracing to this structure to generate 2D images.

When sampling points along a ray during rendering, each point takes the color and density values of the voxel in which it falls, avoiding interpolation between cells to mimic the sharp, pixelated appearance of textures in game. 
To avoid cases where the inside---but not the surface---of a block's voxel grid is sampled during ray tracing, we repeatedly stack face textures within the block, in a pyramidal pattern, to approximate solid objects (Figure~\ref{fig:pyramidal_textures}).

Given a vertex $\*x\in \mathbb{R}^3$ on the solid block grid, we compute a discrete block type, first using an MLP to generate a prediction $\*b_{\mathrm{soft}}\in \mathbb{R}^{M}$ over $M$ block types, then obtaining a onehot vector $\*b_{\mathrm{hard}} \in \left\{e_1, e_2, \dots, e_M\right\}$ using the gumbel softmax function:
\begin{align*}
\begin{aligned}[c]
\*b_{\mathrm{soft}} &= \mathrm{MLP}\left(\*x; \theta_B\right)
\end{aligned}
\qquad
\begin{aligned}[c]
\*b_{\mathrm{hard}} &= \mathrm{gumbel\_softmax}\left(\*b_{\mathrm{soft}}\right),
\end{aligned}
\end{align*}
where $\theta_B$ denotes the parameters of the solid block type MLP. These vectors serve as the elements of $\*B_{\mathrm{soft}}$ and $\*B_{\mathrm{hard}}$ respectively.

Again given the block grid vertex $\*x$, we compute soft and hard air values, where high values correspond to solid blocks and low values approaching $0$ correspond to air.

The soft air grid is computed in the same way as density in previous works, passing the output of an MLP (parameterized by $\theta_A$) through an exponential activation function\footnote{During the backward pass, the exponential function is clipped to avoid exploding gradients: $\sigma_{\mathrm{soft}} = \exp\left(\mathrm{clamp}(y, 15)\right)$.}:
\begin{align*}
\sigma_{\mathrm{soft}} &= \exp\left(\mathrm{MLP}(\mathbf{x}; \theta_{A})\right),
\end{align*}
where $\theta_A$ denotes the parameters of the air block MLP. To quantize the air grid (effectively computing the presence of air blocks), we first use the soft air values to derive the respective probability of an air/solid block appearing at $\*x$:
\begin{align*}
\begin{aligned}[c]
\sigma_{\mathrm{soft}}' &= \sigma_{\mathrm{soft}} - 10
\end{aligned}
\qquad
\begin{aligned}[c]
p_{\mathrm{air}} &= -\sigma_{\mathrm{soft}}'
\end{aligned}
\qquad
\begin{aligned}[c]
p_{\mathrm{solid}} &= \sigma_{\mathrm{soft}}'.
\end{aligned}
\end{align*}

We then use the gumbel softmax function to discretize these predictions, obtaining a density value of $0$ in case of air, and $1$ in case of a solid block:
\begin{align*}
\sigma_{\mathrm{hard}} &= \sum^2\mathrm{gumbel\_softmax}\left(\left[p_{\mathrm{air}}, p_{\mathrm{solid}}\right]\right).
\end{align*}
These values serve as the elements of $\*A_{\mathrm{soft}}$ and $\*A_{\mathrm{hard}}$, respectively.


We use only opaque solid Minecraft blocks (excluding transparent blocks like glass or ice, and porous ones like grass or flowers). 
We use a separate, shallow background MLP, which takes as input a viewing angle, to model color at the end of each ray, allowing the model to learn a low-resolution background texture (effectively projected onto the inside of a sphere).

\subsection{Functional Constraints} \label{sec:functional}

\textbf{Distributional Constraints}
The discrete block grid resulting from the quantized NeRF allows us to optimize \ours{} to produce a level satisfying a target distribution of block types, producing text-guided objects comprising of particular block mixtures. The user sets a target proportion for each block type, and we apply a loss equal to the difference between this target and the actual proportion of (non-air) grid cells in the NeRF's output (after quantization), which we compute by taking the sum over the relevant channel of the discrete onehot block grid and dividing by the size of the grid. Formally, we define the distributional loss as
\begin{align*}
L_D &= \sum_{t\in T} \left|G(t) - P(t)\right|,
\end{align*}
where $t$ is a block type among the set of blocks $T$, $G(t)$ is the target number of occurrences of this block as specified by the user, and $P(t)$ is the number of actual occurrence in the quantized block grid $\*C_{\mathrm{hard}}$.

\begin{figure*}[]
    \begin{subfigure}[b]{.33\textwidth}
        \includegraphics[width=\textwidth]{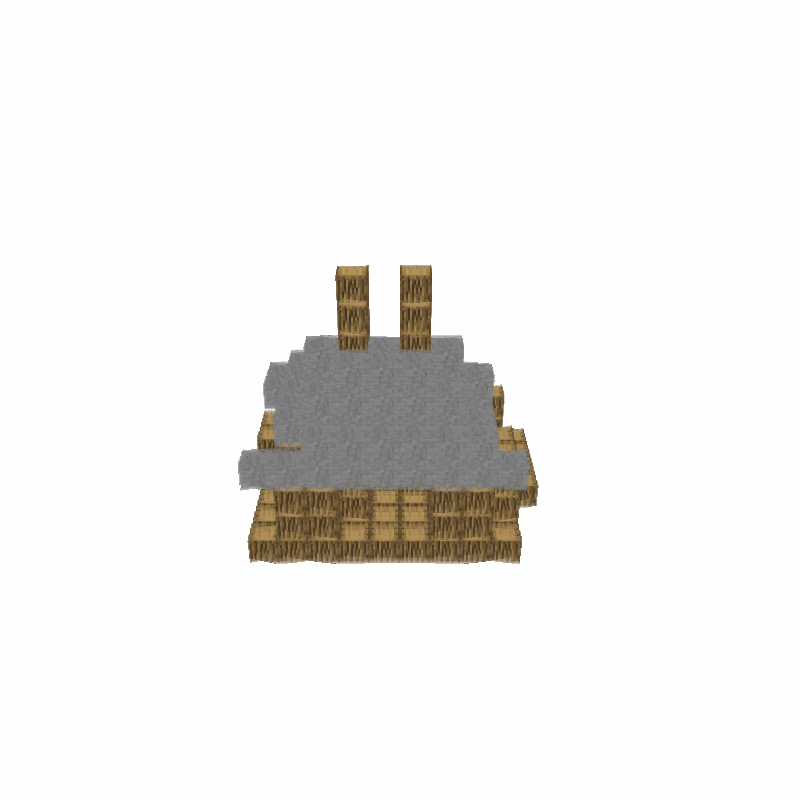}
    \caption*{$N=30$}
    \end{subfigure}
    \begin{subfigure}[b]{.33\textwidth}
        \includegraphics[width=\textwidth]{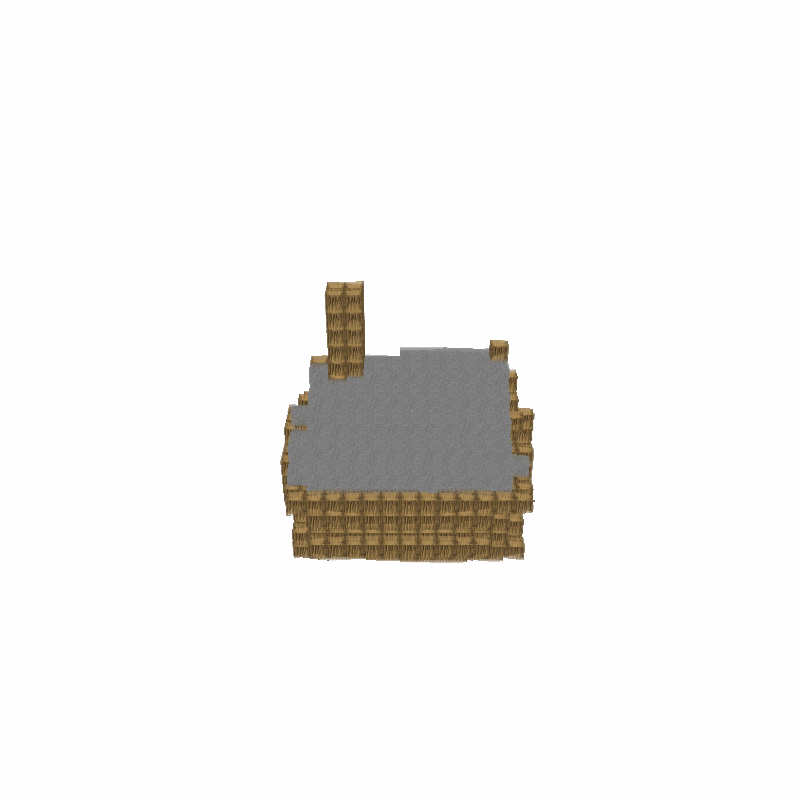}
    \caption*{$N=50$}
    \end{subfigure}
    \begin{subfigure}[b]{.33\textwidth}
        \includegraphics[width=\textwidth]{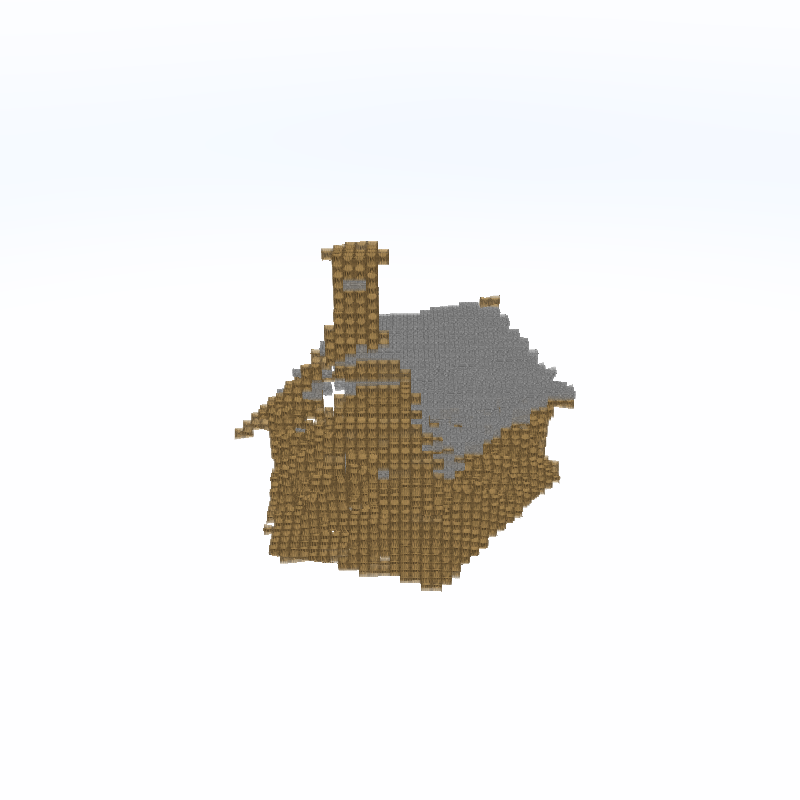}
    \caption*{$N=90$}
    \end{subfigure}
\caption{\textit{``medium medieval home''} with block grids of various widths.}
\end{figure*}

\textbf{Adjacency Constraints} 
We also introduce a loss term corresponding to a penalty or reward incurred whenever a particular configuration of blocks appear in the generated structure. To this end, we construct a convolutional layer that outputs $1$ over any matching patch of blocks, and $0$ everywhere else, then sum the result, yielding the number of occurences of the relevant pattern in $\*C_{\mathrm{hard}}$. We multiply this sum by a user-specified loss coefficient (negative when the pattern is desired, positive when prohibited). Suppose the user wants to apply a loss/reward of $l_p$ to the pattern of blocks $b_0,b_1,\dots,b_{j_p}$ occupying a patch of size $K^3$ (with the number of blocks of interest $j_p \leq K^3$), where each block $b_i$ has relative coordinates $x_i,y_i,z_i$ in the patch. We construct a 3D convolutional weight matrix $W_p$ consisting of the onehot vectors $e_i$ corresponding to each block type, and placing them at position $x_i,y_i,z_i$ in the weight matrix. We apply the resulting convolutional layer, $\textrm{conv}_{W_p}$ to the quantized block grid, subtract $j_p - 1$ from the output, and apply a ReLU activation to obtain the binary pattern-activation tensor. Formally,
\begin{align*}
L_P &=\sum_{p\in P}w_p\sum_{i}^{K^3}\left(\mathrm{ReLU}\left(\textrm{conv}_{W_p}\left(\*C_{\mathrm{hard}}\right) - j_p + 1\right)\right)_i,
\end{align*}
where an inner sum is taken over elements $i$ in the binary activation tensor for pattern $p$.


\begin{table*}
\centering
\caption{\textbf{Fidelity of Neural Renders on COCO and Planet Minecraft} \ours{}'s quantized generations are compared to unquantized generations from an \unconstrained, which can be considered an upper bound for neural renders given their higher resolution. \ours{}'s relative performance increases when moving to a set of domain-relevant prompts.}



\begin{tabular}{r|c|c|c|c}
\toprule
& \multicolumn{2}{c}{COCO} & \multicolumn{2}{c}{Planet Minecraft} \\
 Model & CLIP ViT-B/16 & CLIP ViT-B/32 & CLIP ViT B/16 & CLiP ViT B/32\\
\midrule
Unconstrained NeRF &  \bfseries 61.44 &\bfseries 66.67 &\bfseries 25.17 
&\bfseries 31.29 \\
DreamCraft &   19.74 & 21.05 &  11.56 & 17.01 \\
\midrule
ratio & 0.32 & 0.32 & 0.46 & 0.54 \\ 
\bottomrule
\end{tabular}
%
\label{tab:neural_baseline}
\end{table*}

\begin{table}
\centering
\caption{\textbf{Fidelity of In-Game Renders on Planet Minecraft} \ours{}, which learns to use discrete blocks during training, outperforms a baseline whose output is discretized only after training.}
\begin{tabular}{c|r|r}
\toprule
Model &  CLIP ViT-B/16 & CLIP ViT-B/32 \\
\midrule
Unconstrained NeRF &  2.72 & 2.72  \\
DreamCraft &  \bfseries 5.44 & \bfseries 6.12  \\
\bottomrule
\end{tabular}

\label{tab:planet_mc_ingame}
\end{table}

\section{Experiments}
\textbf{Baselines} We compare the performance of \ours{}, our quantized NeRF which learns to arrange in-game assets \textit{during training}, to \textbf{\unconstrained{}}, a baseline that maps the continuous outputs to game assets \textit{after training}. \unconstrained{} trains a text-guided NeRF and then maps its output to Minecraft blocks using nearest neighbors. For text-guidance, we use a variant of Emu~\cite{dai2023emu} trained only on Shutterstock, a model using text-conditioned latent diffusion to generate images. The nearest RGB value to each Minecraft block is determined by taking the average color over each of the (normally repeated) $16\times 16$ textures covering each of its 6 faces. We define a width-$N$ grid over the 3D output space of the unconstrained model and query the RGB and density values at the center of each cell in the grid. We calculate the $L_2$ distance between each centerpoint and average Minecraft block color, mapping each cell to the closest block. We then select a density threshold $s=10$, and place air blocks wherever $\sigma < s$. 

\textbf{Ablations} We study different quantization schemes in order to understand what is the best way to map the continuous outputs of the air and solid block MLPs into discrete grids of Minecraft blocks. The output of either MLP can be passed through the gumbel softmax function to produce a discrete grid of air or solid blocks (see Figure~\ref{fig:overview}). If these values are not discretized \ie{} $\alpha < 1$ or $\beta <1$, meaning that $\*b_{\mathrm{soft}}$ or $\sigma_{\mathrm{soft}}$ are used instead of their ``hard'' counterparts, then the resulting voxel grid can include solid blocks interpolated with air blocks (\ie{} semi-transparent) or with one another (\ie{} multi-texture). We also experiment with linearly annealing these values from their soft to hard counterparts over the course of training.

\textbf{Evaluation Datasets} We evaluate our method on both generic and domain-specific text prompts, using the \textbf{COCO}~\cite{lin2014microsoft} and \textbf{Planet Minecraft}~\cite{planet-minecraft} datasets, respectively. For COCO, we use the same 153 prompts as in prior text-to-3D works~\cite{mohammad2022clip,poole2022dreamfusion,lee2022understanding}. For Planet Minecraft, we take the names of the top $150$ most downloaded assets uploaded by users in $2016$ under the ``Maps" category. Some examples include ``a desperate and lonely wizards tower pmc chunk challenge entry lore'', ``mario kartgba bowsers castle 2'', and ``icarly set and nickelodeon studio''. A full list of prompts can be found in Section~\ref{sec:planet_mc_dataset}.

\textbf{Evaluation Metrics} To quantitatively evaluate the performance of our model, we measure its fidelity using \textbf{R-precision}. More specifically, we query other pre-trained joint text-image encoders, namely CLIP ViT-B/16 and CLIP ViT-B/32~\cite{radford2021learning}, and test whether they can recognize the caption responsible for a given NeRF rendering from a set of distractors (other, randomly selected captions from the dataset). For each caption, we repeat the process for 5 different test images and average the result.



\begin{table}
\centering
\caption{\textbf{Fidelity of Different Quantization Schemes for Planet Minecraft}. Using a hard block type and soft block density achieves the best performance.}
\begin{tabularx}{\columnwidth}{X|X|X|X|X|X}
\toprule
    &  & \multicolumn{2}{r}{CLIP ViT-B/16} & \multicolumn{2}{r}{CLIP ViT-B/32} \\
    &  & RGB & depth & RGB & depth \\
 block type & block density &   &  &  &  \\
 \midrule
 \multirow[c]{3}{*}{anneal} & anneal & 7.73 & 5.07 & 9.07 & 5.73 \\
   & hard &  2.67 & 1.20 & 5.60 & 2.00 \\
   & soft &  7.33 & 7.33 & 10.40 & 5.20 \\
   \midrule
  \multirow[c]{3}{*}{hard} & anneal &  8.53 & 6.00 & 8.40 & 7.33 \\
   & hard &  2.27 & 1.33 & 2.80 & 1.87 \\
   & soft & \bfseries 10.40 &\bfseries 8.93 &\bfseries 13.20 &\bfseries 8.40 \\
   \midrule
  \multirow[c]{3}{*}{soft} & anneal &  8.40 & 3.60 & 10.00 & 4.80 \\
   & hard &  4.00 & 0.80 & 7.60 & 2.80 \\
   & soft &  7.87 & 6.53 & 11.07 & 5.47 \\
   \bottomrule
\end{tabularx}
%
\label{tab:planet_mc_quant}
\end{table}

\begin{table}[]
\centering
\caption{Fidelity of generated structures given different quantization schemes for block density and type. COCO dataset, neural renders. Both RGB screenshots and depth-only greyscale heatmaps of the generated 3D structure are evaluated. A combination of fully discrete block types and ``soft'' block density leads to best performance in terms of depth (i.e. structure topology), while soft block types and density during training lead to the best RGB images. R-precision averaged over 5 poses for each experiment.}
\begin{tabularx}{\linewidth}{X|X|X|X|X|X}
\toprule
   &  & \multicolumn{2}{c}{CLIP ViT-B/16} & \multicolumn{2}{c}{CLIP ViT-B/32} \\
   \hline
  &   & RGB & depth & RGB & depth \\
 block type & block density  &  &  &  &  \\
\midrule
 \multirow[c]{3}{*}{anneal} & anneal  & 12.68 & 4.31 & 11.37 & 3.01 \\
  & hard  & 5.36 & 0.65 & 6.80 & 0.92 \\
  & soft  & 22.88 & 8.89 & 22.88 & 8.24 \\
  \midrule
\multirow[c]{3}{*}{hard} & anneal  & 12.68 & 5.88 & 14.12 & 4.71 \\
 & hard  & 4.05 & 2.22 & 4.97 & 1.18 \\
  & soft  & 19.61 & \bfseries 12.03 & 21.83 & \bfseries 11.11 \\
  \midrule
 \multirow[c]{3}{*}{soft} & anneal  & 17.65 & 4.84 & 15.69 & 4.44 \\
  & hard  & 9.80 & 1.05 & 7.32 & 0.92 \\
  & soft & \bfseries 24.31 & 8.76 & \bfseries 24.97 & 7.32 \\
  \bottomrule
\end{tabularx}
%
\label{tab:coco_quant}
\end{table}

\subsection{Quality of the Generations}

In Figures~\ref{fig:planet_mc_unconstrained} and \ref{fig:coco_unconstrained}, we can see that, using only Minecraft blocks, our model produces structures that are visually similar to those of the \unconstrained{}, for both generic and domain-specific text prompts. 
Note that the \unconstrained{} model is a strong upper bound because it has a continuous and thus much larger output space (i.e., higher resolution) than our discretized \ours{} model.
\ours{}’s relative performance increases when moving to a set of
domain-relevant prompts.

In Table~\ref{tab:neural_baseline}, we compare the fidelity of \ours{} with that of the \unconstrained{} on COCO and Planet Minecraft. Note that here, the generations are evaluated using the renders from the neural ray tracing engine rather than in-game renders. As expected, limiting the NeRF's output to a specific set of discretely assembled blocks drastically reduces its space of generations. This is reflected in \ours{}'s lower fidelity with respect to the \unconstrained{}, when generating objects from both the COCO dataset and Planet Minecraft datasets. However, the performance gap between \ours{} and the \unconstrained{} is reduced when moving from the generic COCO dataset to the domain-specific Planet Minecraft dataset. This suggests that despite its restricted output space, \ours{} is particularly effective at generating high-quality structures when the input prompts are relevant to the (discrete)
domain at hand. 



In Table~\ref{tab:planet_mc_ingame}, we evaluate the R-precision using 2D captures of generated Minecraft block layouts in the game engine itself. Note that here, the generations are evaluated using the in-game renders rather than the neural renders. For game design applications, in-game render fidelity is more relevant than neural render fidelity, so this is ultimately the metric we care about in our study. In this case, the fidelity of the Unconstrained NeRF is lower than that of \ours{} for both COCO and Planet Minecraft. This indicates that post-processing the output of a NeRF in order to discretize it using nearest-neighbor leads to worse results than learning to use discrete blocks during the generation process.  
In Figure~\ref{fig:postprocess_baseline}, we see that mapping a discrete set of grid vertices to nearest neighbor block types via average color leads to sub-optimal results that are particularly bad at maintaining consistency in terms of texture and color.
This result demonstrates the difficulty of translating unconstrained generations to a constrained repertoire of domain-specific assets.
We conclude that by incorporating these game assets in the learning process, we can generate more faithful in-game structures.

\subsection{Quantization Schemes}

In Table~\ref{tab:planet_mc_quant}, we compare the effect of applying soft, hard, and annealed quantization schemes to solid and air blocks. Maintaining \textit{soft air blocks} (continuous-valued block transparency), in combination with \textit{hard solid blocks} (discrete block types), leads to the highest R-precision at test time.
This suggests that learning the topology (in contrast to the color/texture) of a generated structure is a sensitive process in which relaxing the quantization scheme (and presumably simplifying the loss landscape) is crucial.

\begin{figure}[ht]
\setlength{\tabcolsep}{1pt}
\centering
\par%
\begin{tabular}{cccc}
& soft blocks & anneal & hard blocks\\
\begin{sideways}hard air\end{sideways}& \includegraphics[width=.3\linewidth]{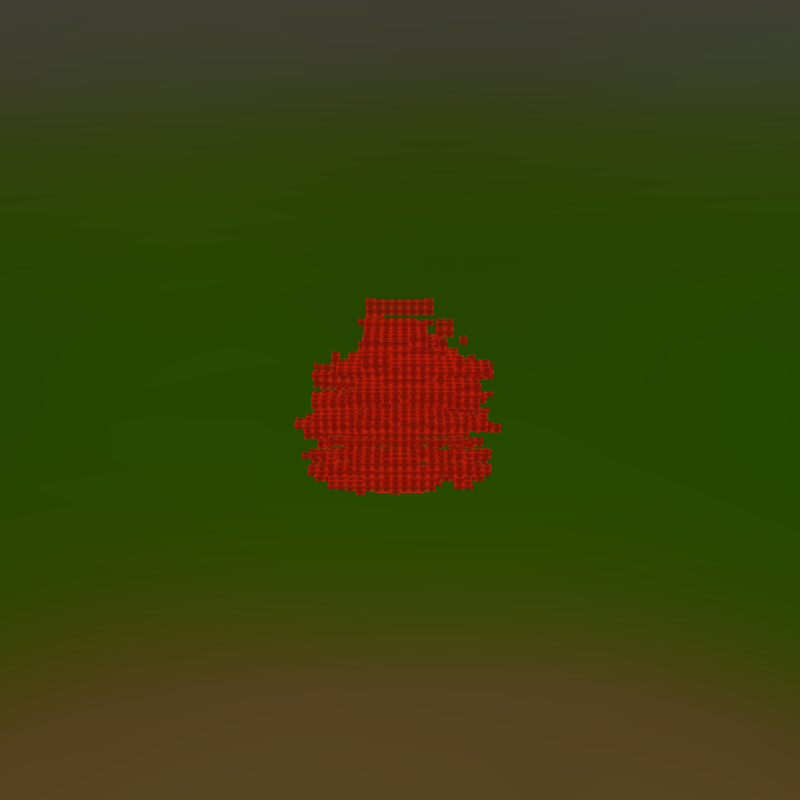} & \includegraphics[width=.3\linewidth]{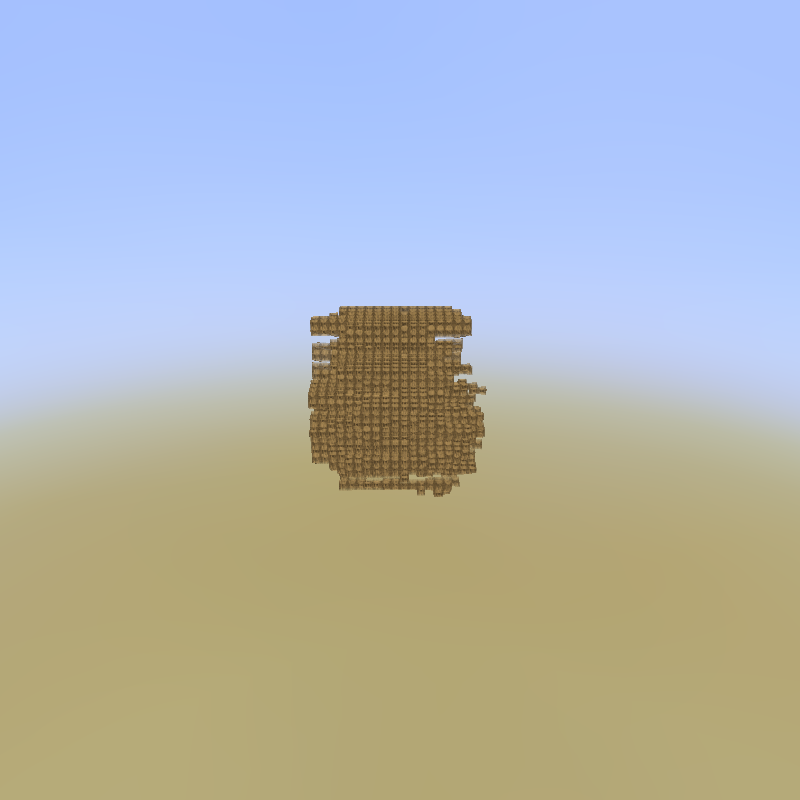} & \includegraphics[width=.3\linewidth]{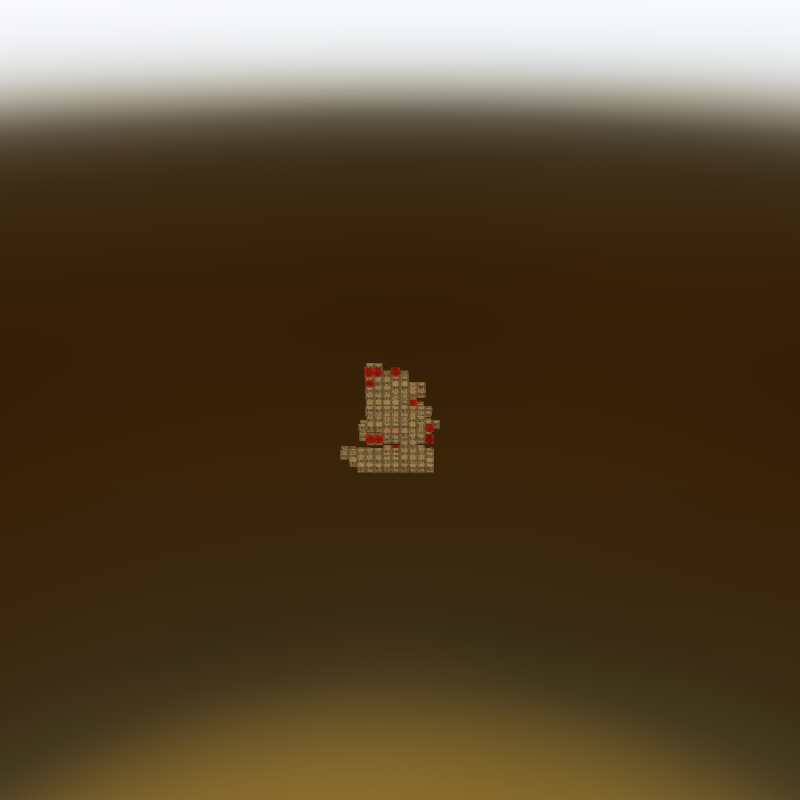}\\
\begin{sideways}anneal \end{sideways}& \includegraphics[width=.3\linewidth]{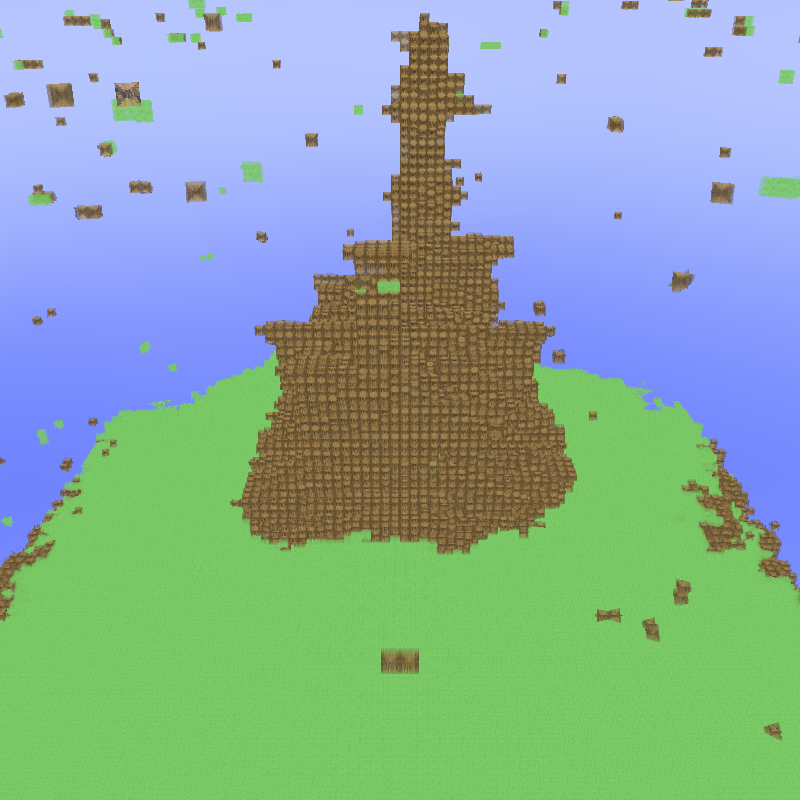} & \includegraphics[width=.3\linewidth]{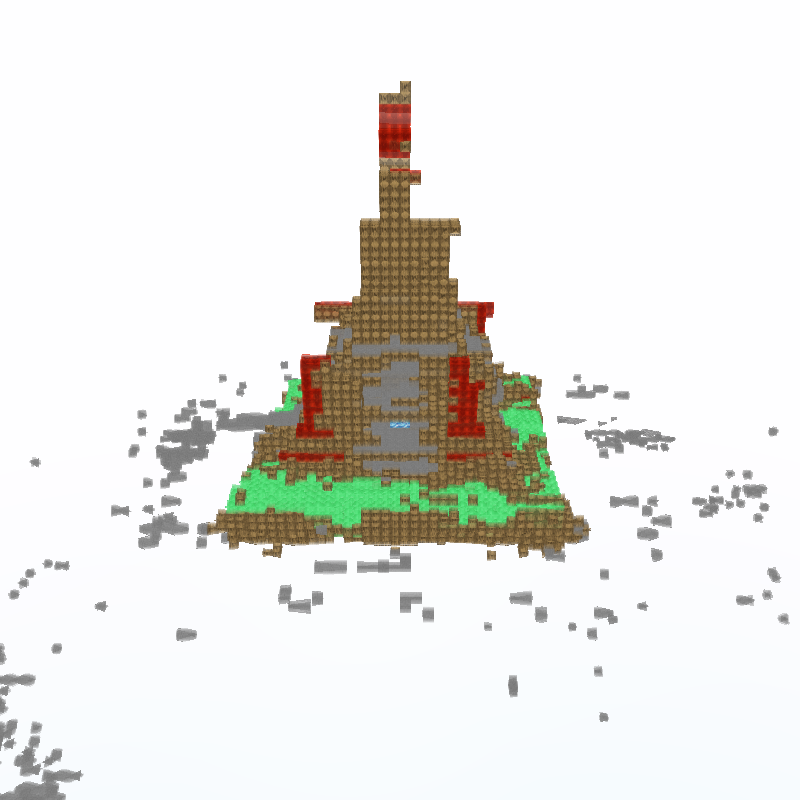} & \includegraphics[width=.3\linewidth]{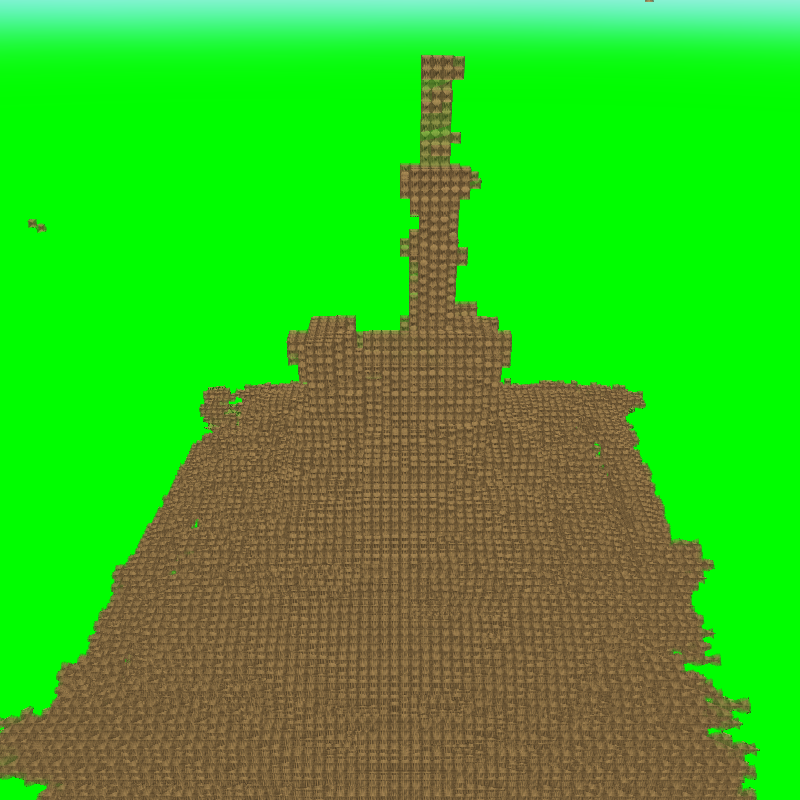}\\
\begin{sideways}soft air \end{sideways}& \includegraphics[width=.3\linewidth]{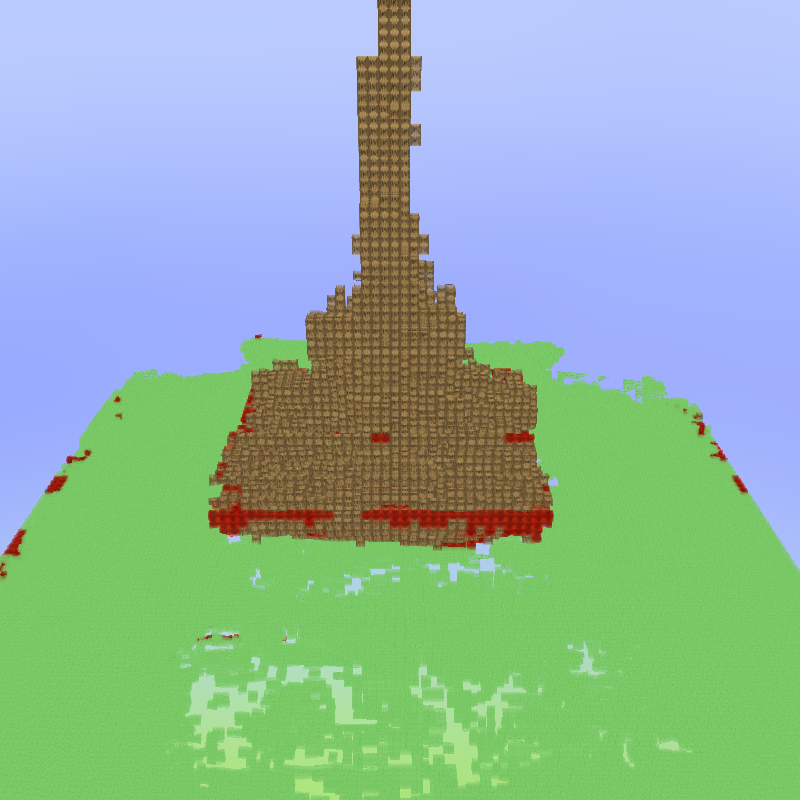} & \includegraphics[width=.3\linewidth]{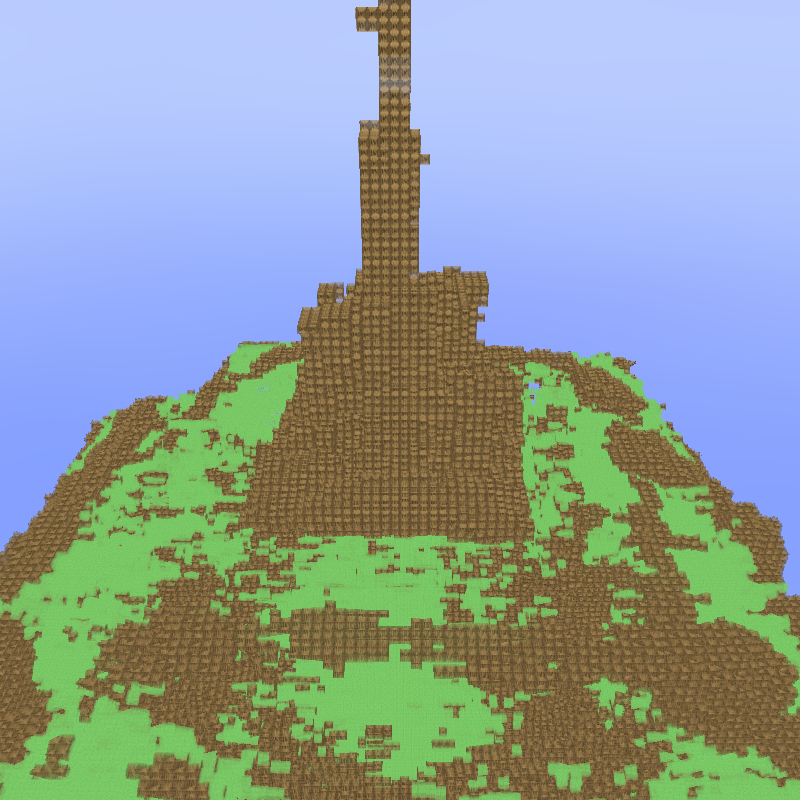} & \includegraphics[width=.3\linewidth]{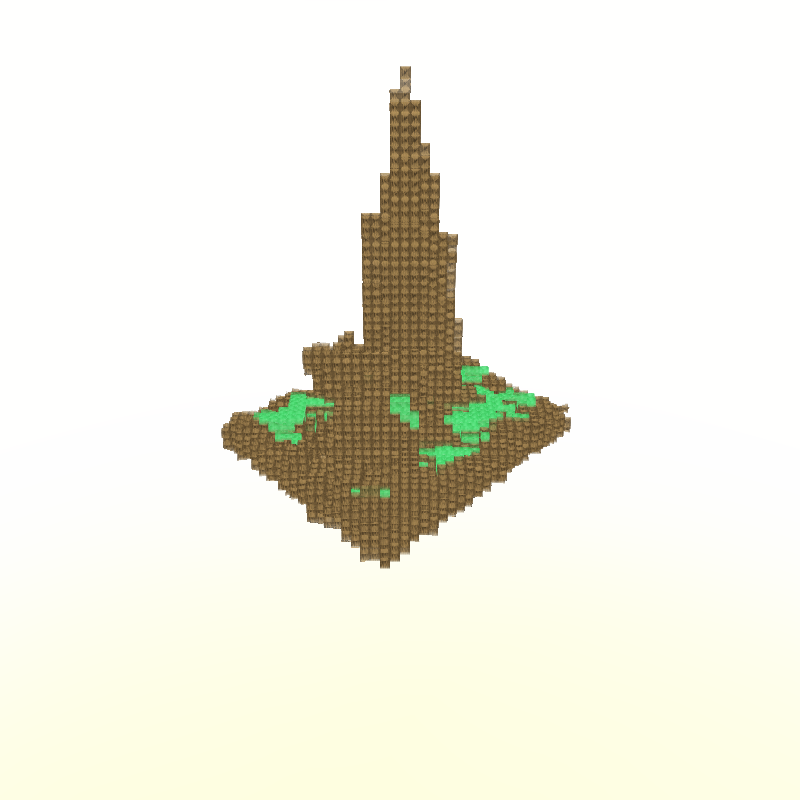}\\
\end{tabular}
\caption{\textbf{Qualitative Effect of Different Quantization Schemes} for \textit{``dwarven entrance''}. The best results obtained using hard block types and soft air blocks. Meaningful representations are learned only when the discreteness of air blocks (or by analogy, the density or topology of the generated structure) is relaxed or annealed throughout training (bottom two rows).}.%
\label{fig:dwarf_quant}
\end{figure}

Forcing block density to be discrete throughout training leads to poor performance (as can be seen in the first row of Figure~\ref{fig:dwarf_quant}), with the poorest performance coming from models in which both block type and density are fully discrete. 
This may be because of noisier learning dynamics resulting from the quantized output space of the model.  

Conversely, using soft block density can lead to situations in which apparently solid surfaces are emulated by layering a number of semi-transperent blocks. At test time, when rendering the fully discrete block grid, such surfaces can suddenly be culled from the image, as none of the individual blocks of which they consist are enough to result in a ``solid'' binary output after quantization.

\section{Block Grid Resolution}

\begin{table}
\centering
\caption{Fidelity of generated structures given block grids of varying resolutions. Planet Minecraft dataset. Neural renders. Performance increases with the resolution of the block grid.}
\begin{tabularx}{\columnwidth}{X|X|X|X|X}
\toprule
 &  \multicolumn{2}{c}{CLIP ViT-B/16} & \multicolumn{2}{c}{CLIP ViT-B/32} \\
\hline
 &   RGB & depth & RGB & depth \\
 block grid &  &  &  &  \\
 \midrule
20 & 5.47 & 2.40 & 6.00 & 2.93 \\
40 & 6.53 & 6.13 & 8.67 & 6.93 \\
60 & 7.33 & 6.80 & 9.33 & 4.40 \\
80 & 11.73 & 8.27 & 12.27 & 6.67 \\
100 & \bfseries 10.80 &\bfseries 9.07 &\bfseries 13.33 & 8.67 \\
\bottomrule
\end{tabularx}
\label{tab:planet_mc_grid}
\end{table}

\begin{figure*}
\begin{subfigure}[b]{.49\textwidth}
\centering
\begin{subfigure}[t]{.32\textwidth}
\includegraphics[width=\textwidth]{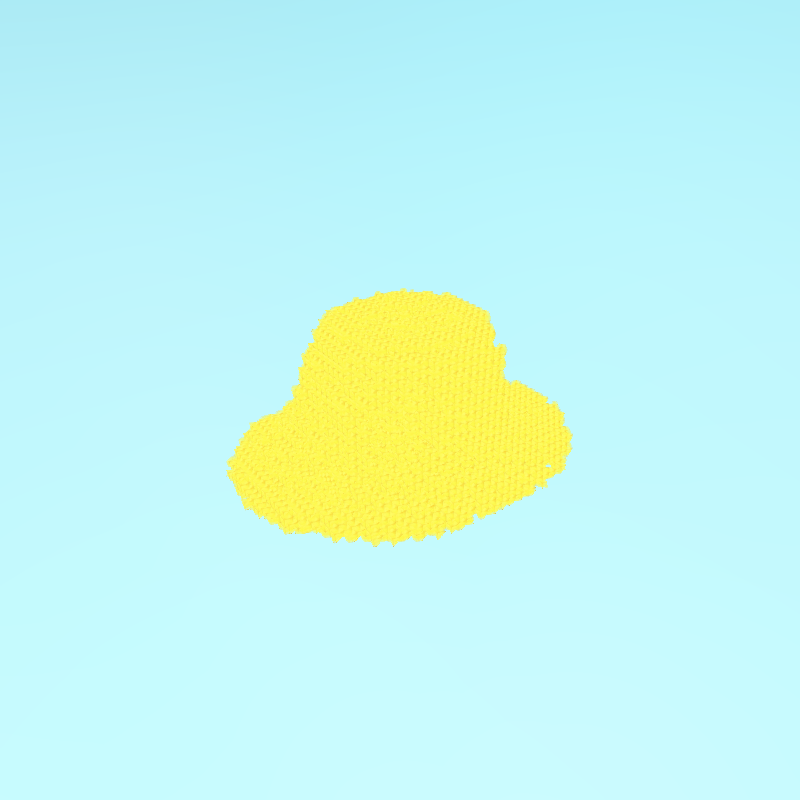}
\subcaption*{$n(\text{gold}) = 1.0$}
\end{subfigure}
\begin{subfigure}[t]{.32\textwidth}
\includegraphics[width=\textwidth]{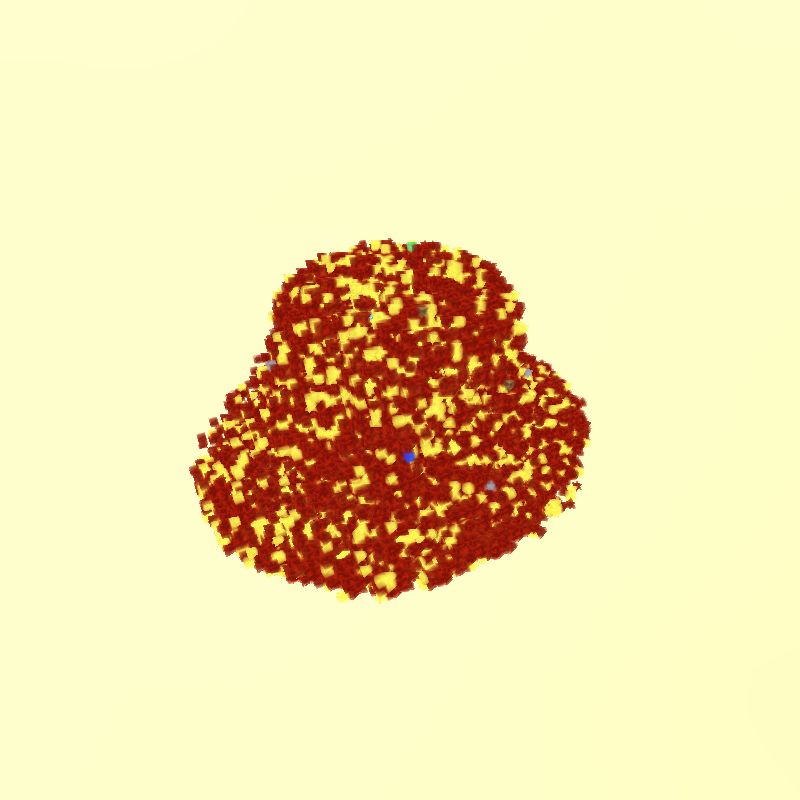}
\subcaption*{$n(\text{redstone}) = n(\text{gold})$}
\end{subfigure}
\begin{subfigure}[t]{.32\textwidth}
\includegraphics[width=\textwidth]{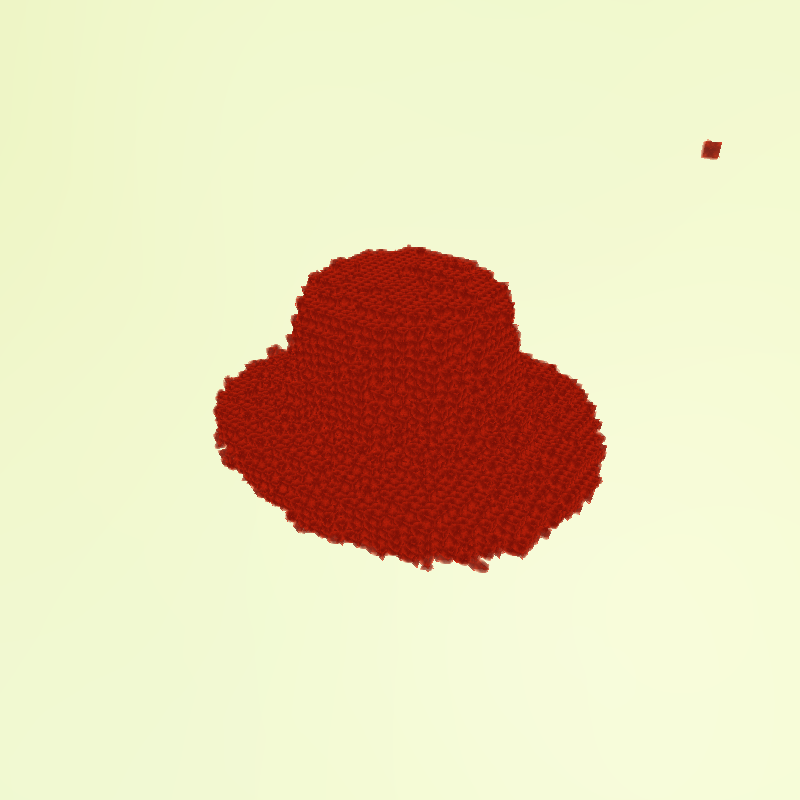}
\subcaption*{$n(\text{redstone}) = 1.0$}
\end{subfigure}
\caption{\textbf{Distributional Constraints.} \textit{``a stylish hat''} jointly optimized to satisfy a given target distribution over the block types.}
\label{fig:distr_con}
\end{subfigure}
\begin{subfigure}[b]{.49\textwidth}
\centering
\begin{subfigure}[b]{.32\textwidth}
\includegraphics[width=\textwidth]{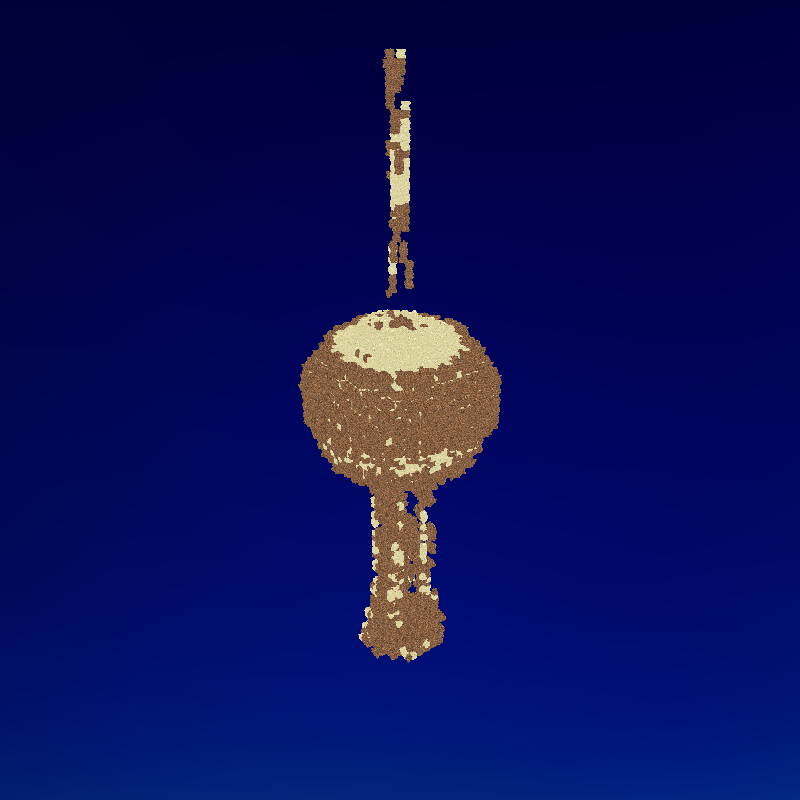}
\subcaption*{$w=0.25$}
\end{subfigure}
\begin{subfigure}[b]{.32\textwidth}
\includegraphics[width=\textwidth]{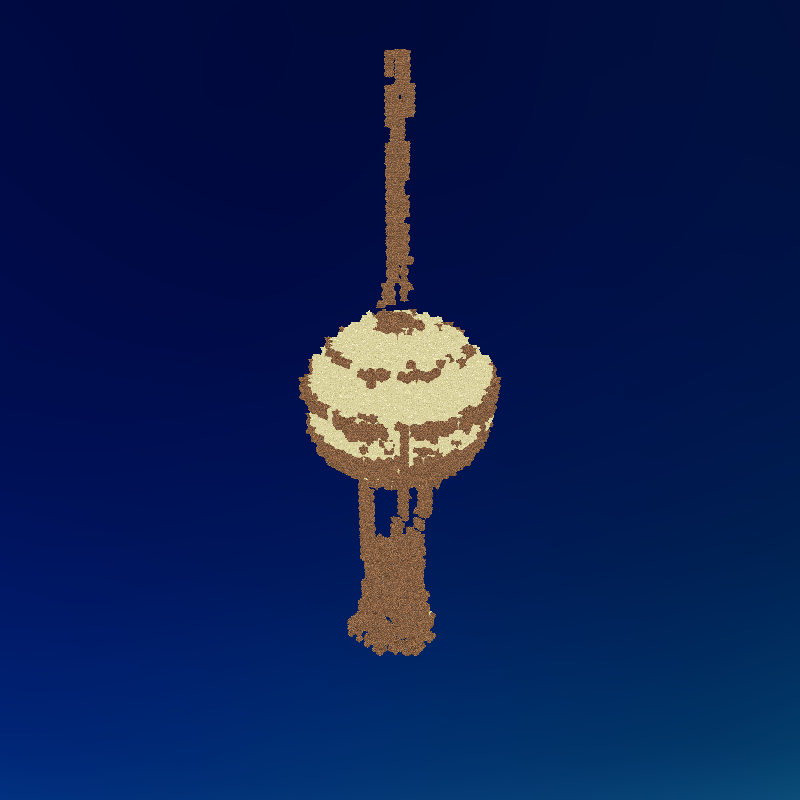}
\subcaption*{$w=1.75$}
\end{subfigure}
\begin{subfigure}[b]{.32\textwidth}
\includegraphics[width=\textwidth]{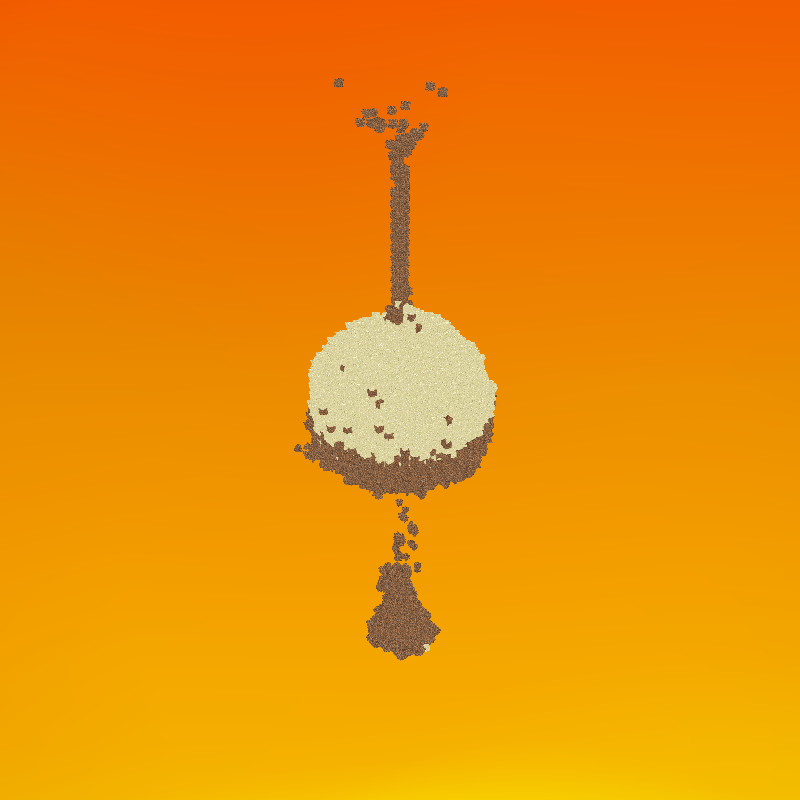}
\subcaption*{$w=2$}
\end{subfigure}
\caption{\textbf{Adjacency Constraints.} \textit{``space needle accurate''} with a penalty for floating sand, weighted more heavily from left to right. The block set is limited to sand (light tan) and dirt (brown).}
\label{fig:adj_con}
\end{subfigure}
\caption{Examples of incorporation of functional constraints via auxiliary loss terms.}
\end{figure*}

We experiment with the resolution of the block grid, learning a $N\times N\times N$-block representation of text prompts with $N\in\{10, 20,\dots, 100\}$.

In Table~\ref{tab:planet_mc_grid}, we see that increasing block grid resolution leads to an increase in R-precision. We note that the more blocks in the grid, the closer each block comes to being represented by only a single pixel in each 2D render of the 3D block layout. In other words, we can expect the output of these higher-resolution quantized NeRFs to approximate that of their unconstrained counterparts with increasing accuracy. By analogy with visual art, we can say that the model uses blocks in an increasingly pointillistic fashion.

\subsection{Functional Constraints}

In Figure~\ref{fig:distr_con}, we illustrate the effect of adding distributional constraints to a prompt asking for ``a stylish hat''. We can see that the model produces a similar structure using either entirely gold, redstone, or an even split of both when adding the distributional loss term.

In Figure~\ref{fig:adj_con}, we illustrate the effect of adding an adjacency constraint prohibiting sand from ``floating'', i.e. being placed directly above an air tile, and ask for ``space needle accurate'' (a Planet Minecraft prompt). As the weight of the adjacency loss term is increased, the use of sand blocks becomes restricted to the central ``bulb'' of the tower, where it is more likely to be supported by a the circular base of dirt blocks. When the adjacency loss is weighted less heavily, sand is often incorporated into the underside of the bulb (where it is unsupported and will fall to the ground in-game).

These experiments demonstrate that functional constraints can be easily integrated with our text-to-3D model to create controllable Minecraft structures that obey both high-level and low-level user specifications and can be rendered in-game. 


\section{Conclusion}

In this work, we develop a new approach for generating functional game environments in Minecraft from free-form text descriptions. \ours{} quantizes the output of a text-to-3D NeRF to predict discrete block types which are then mapped to game assets (\ie{} voxel-grids corresponding to in-game blocks). This allows the NeRF to use the game assets to represent the content described by a text prompt. We demonstrate that our approach has higher fidelity to the text prompt than a baseline that discretizes the output of an \unconstrained{} \textit{after} learning. \ours{} is, to our knowledge, the first generator capable of generating diverse, functional, and controllable 3D game environments directly from free-form text. Since our model can adapt to the unique appearance of user-supplied modular game assets to produce environments with high-level aesthetic properties, it may be particularly useful for game designers working in new domains that don't yet have have large datasets of game layouts. 

One limitation of \ours{} is that it takes a few hours to generate a single structure. However, it could benefit from recent and future speed improvements in NeRFs~\cite{wang2023f,guo2022neural,sun2022direct,zhang2020video}. Another promising direction for future work is to model lightning and shadow, in addition to color and density, which could be achieved using an auxiliary loss to model the in-game lightning effects.

While we focus on MineCraft, \ours{} could be extended beyond cube-based environments, to any 3D environment involving discrete assets that can be approximated by voxel grids. It could also incorporate more complex functional constraints, assuming these could be implemented to be differentiable. For example, one could compute the path length between key blocks (e.g. a player spawn block or a treasure chest) using convolutions, and use the difference between the target achieved path lengths as part of the loss, similar to the reward used in \citet{khalifa2020pcgrl,earle2021learning}. It may also be beneficial---especially where functioinal constraints cannot be made differentiable---to use RL methods to approximate the gradient from such additional functional scores. Vis-a-vis embodied player agents, environments could be generated to result in certain rewards, dynamics, regret or learnability with respect to these agents.



\bibliographystyle{ACM-Reference-Format}
\bibliography{bib}


\begin{thebibliography}{93}


\ifx \showCODEN    \undefined \def \showCODEN     #1{\unskip}     \fi
\ifx \showDOI      \undefined \def \showDOI       #1{#1}\fi
\ifx \showISBNx    \undefined \def \showISBNx     #1{\unskip}     \fi
\ifx \showISBNxiii \undefined \def \showISBNxiii  #1{\unskip}     \fi
\ifx \showISSN     \undefined \def \showISSN      #1{\unskip}     \fi
\ifx \showLCCN     \undefined \def \showLCCN      #1{\unskip}     \fi
\ifx \shownote     \undefined \def \shownote      #1{#1}          \fi
\ifx \showarticletitle \undefined \def \showarticletitle #1{#1}   \fi
\ifx \showURL      \undefined \def \showURL       {\relax}        \fi
\providecommand\bibfield[2]{#2}
\providecommand\bibinfo[2]{#2}
\providecommand\natexlab[1]{#1}
\providecommand\showeprint[2][]{arXiv:#2}

\bibitem[Alvarez et~al\mbox{.}(2018)]%
        {alvarez2018assessing}
\bibfield{author}{\bibinfo{person}{Alberto Alvarez}, \bibinfo{person}{Steve
  Dahlskog}, \bibinfo{person}{Jose Font}, \bibinfo{person}{Johan Holmberg},
  {and} \bibinfo{person}{Simon Johansson}.} \bibinfo{year}{2018}\natexlab{}.
\newblock \showarticletitle{Assessing aesthetic criteria in the evolutionary
  dungeon designer}. In \bibinfo{booktitle}{\emph{Proceedings of the 13th
  International Conference on the Foundations of Digital Games}}.
  \bibinfo{pages}{1--4}.
\newblock


\bibitem[Awiszus et~al\mbox{.}(2021)]%
        {awiszus2021world}
\bibfield{author}{\bibinfo{person}{Maren Awiszus}, \bibinfo{person}{Frederik
  Schubert}, {and} \bibinfo{person}{Bodo Rosenhahn}.}
  \bibinfo{year}{2021}\natexlab{}.
\newblock \showarticletitle{World-gan: a generative model for minecraft
  worlds}. In \bibinfo{booktitle}{\emph{2021 IEEE Conference on Games (CoG)}}.
  IEEE, \bibinfo{pages}{1--8}.
\newblock


\bibitem[Baker et~al\mbox{.}(2022)]%
        {baker2022video}
\bibfield{author}{\bibinfo{person}{Bowen Baker}, \bibinfo{person}{Ilge Akkaya},
  \bibinfo{person}{Peter Zhokov}, \bibinfo{person}{Joost Huizinga},
  \bibinfo{person}{Jie Tang}, \bibinfo{person}{Adrien Ecoffet},
  \bibinfo{person}{Brandon Houghton}, \bibinfo{person}{Raul Sampedro}, {and}
  \bibinfo{person}{Jeff Clune}.} \bibinfo{year}{2022}\natexlab{}.
\newblock \showarticletitle{Video pretraining (vpt): Learning to act by
  watching unlabeled online videos}.
\newblock \bibinfo{journal}{\emph{Advances in Neural Information Processing
  Systems}}  \bibinfo{volume}{35} (\bibinfo{year}{2022}),
  \bibinfo{pages}{24639--24654}.
\newblock


\bibitem[Bontrager and Togelius(2021)]%
        {bontrager2021learning}
\bibfield{author}{\bibinfo{person}{Philip Bontrager} {and}
  \bibinfo{person}{Julian Togelius}.} \bibinfo{year}{2021}\natexlab{}.
\newblock \showarticletitle{Learning to generate levels from nothing}. In
  \bibinfo{booktitle}{\emph{2021 IEEE Conference on Games (CoG)}}. IEEE,
  \bibinfo{pages}{1--8}.
\newblock


\bibitem[Brewer(2017)]%
        {brewer2017computerized}
\bibfield{author}{\bibinfo{person}{Nathan Brewer}.}
  \bibinfo{year}{2017}\natexlab{}.
\newblock \showarticletitle{Computerized Dungeons and Randomly Generated
  Worlds: From Rogue to Minecraft [Scanning Our Past]}.
\newblock \bibinfo{journal}{\emph{Proc. IEEE}} \bibinfo{volume}{105},
  \bibinfo{number}{5} (\bibinfo{year}{2017}), \bibinfo{pages}{970--977}.
\newblock


\bibitem[Brocchini et~al\mbox{.}(2022)]%
        {brocchini2022monster}
\bibfield{author}{\bibinfo{person}{Michele Brocchini}, \bibinfo{person}{Marco
  Mameli}, \bibinfo{person}{Emanuele Balloni}, \bibinfo{person}{Laura~Della
  Sciucca}, \bibinfo{person}{Luca Rossi}, \bibinfo{person}{Marina Paolanti},
  \bibinfo{person}{Emanuele Frontoni}, {and} \bibinfo{person}{Primo
  Zingaretti}.} \bibinfo{year}{2022}\natexlab{}.
\newblock \showarticletitle{MONstEr: A Deep Learning-Based System for the
  Automatic Generation of Gaming Assets}. In \bibinfo{booktitle}{\emph{Image
  Analysis and Processing. ICIAP 2022 Workshops: ICIAP International Workshops,
  Lecce, Italy, May 23--27, 2022, Revised Selected Papers, Part I}}. Springer,
  \bibinfo{pages}{280--290}.
\newblock


\bibitem[Bruce et~al\mbox{.}(2024)]%
        {bruce2024genie}
\bibfield{author}{\bibinfo{person}{Jake Bruce}, \bibinfo{person}{Michael
  Dennis}, \bibinfo{person}{Ashley Edwards}, \bibinfo{person}{Jack
  Parker-Holder}, \bibinfo{person}{Yuge Shi}, \bibinfo{person}{Edward Hughes},
  \bibinfo{person}{Matthew Lai}, \bibinfo{person}{Aditi Mavalankar},
  \bibinfo{person}{Richie Steigerwald}, \bibinfo{person}{Chris Apps},
  {et~al\mbox{.}}} \bibinfo{year}{2024}\natexlab{}.
\newblock \showarticletitle{Genie: Generative Interactive Environments}.
\newblock \bibinfo{journal}{\emph{arXiv preprint arXiv:2402.15391}}
  (\bibinfo{year}{2024}).
\newblock


\bibitem[Canossa and Smith(2015)]%
        {canossa2015towards}
\bibfield{author}{\bibinfo{person}{Alessandro Canossa} {and}
  \bibinfo{person}{Gillian Smith}.} \bibinfo{year}{2015}\natexlab{}.
\newblock \showarticletitle{Towards a procedural evaluation technique: Metrics
  for level design}. In \bibinfo{booktitle}{\emph{The 10th International
  Conference on the Foundations of Digital Games}}. sn, \bibinfo{pages}{8}.
\newblock


\bibitem[Chen et~al\mbox{.}(2022)]%
        {chen2022tensorf}
\bibfield{author}{\bibinfo{person}{Anpei Chen}, \bibinfo{person}{Zexiang Xu},
  \bibinfo{person}{Andreas Geiger}, \bibinfo{person}{Jingyi Yu}, {and}
  \bibinfo{person}{Hao Su}.} \bibinfo{year}{2022}\natexlab{}.
\newblock \showarticletitle{Tensorf: Tensorial radiance fields}. In
  \bibinfo{booktitle}{\emph{Computer Vision--ECCV 2022: 17th European
  Conference, Tel Aviv, Israel, October 23--27, 2022, Proceedings, Part
  XXXII}}. Springer, \bibinfo{pages}{333--350}.
\newblock


\bibitem[Dahlskog and Togelius(2014)]%
        {dahlskog2014procedural}
\bibfield{author}{\bibinfo{person}{Steve Dahlskog} {and}
  \bibinfo{person}{Julian Togelius}.} \bibinfo{year}{2014}\natexlab{}.
\newblock \showarticletitle{Procedural content generation using patterns as
  objectives}. In \bibinfo{booktitle}{\emph{Applications of Evolutionary
  Computation: 17th European Conference, EvoApplications 2014, Granada, Spain,
  April 23-25, 2014, Revised Selected Papers 17}}. Springer,
  \bibinfo{pages}{325--336}.
\newblock


\bibitem[Dai et~al\mbox{.}(2023)]%
        {dai2023emu}
\bibfield{author}{\bibinfo{person}{Xiaoliang Dai}, \bibinfo{person}{Ji Hou},
  \bibinfo{person}{Chih-Yao Ma}, \bibinfo{person}{Sam Tsai},
  \bibinfo{person}{Jialiang Wang}, \bibinfo{person}{Rui Wang},
  \bibinfo{person}{Peizhao Zhang}, \bibinfo{person}{Simon Vandenhende},
  \bibinfo{person}{Xiaofang Wang}, \bibinfo{person}{Abhimanyu Dubey},
  {et~al\mbox{.}}} \bibinfo{year}{2023}\natexlab{}.
\newblock \showarticletitle{Emu: Enhancing image generation models using
  photogenic needles in a haystack}.
\newblock \bibinfo{journal}{\emph{arXiv preprint arXiv:2309.15807}}
  (\bibinfo{year}{2023}).
\newblock


\bibitem[Dennis et~al\mbox{.}(2020)]%
        {dennis2020emergent}
\bibfield{author}{\bibinfo{person}{Michael Dennis}, \bibinfo{person}{Natasha
  Jaques}, \bibinfo{person}{Eugene Vinitsky}, \bibinfo{person}{Alexandre
  Bayen}, \bibinfo{person}{Stuart Russell}, \bibinfo{person}{Andrew Critch},
  {and} \bibinfo{person}{Sergey Levine}.} \bibinfo{year}{2020}\natexlab{}.
\newblock \showarticletitle{Emergent complexity and zero-shot transfer via
  unsupervised environment design}.
\newblock \bibinfo{journal}{\emph{Advances in neural information processing
  systems}}  \bibinfo{volume}{33} (\bibinfo{year}{2020}),
  \bibinfo{pages}{13049--13061}.
\newblock


\bibitem[Dieterich(2017)]%
        {dieterich2017using}
\bibfield{author}{\bibinfo{person}{Robert~Ota Dieterich}.}
  \bibinfo{year}{2017}\natexlab{}.
\newblock \emph{\bibinfo{title}{Using Proof-Of-Concept Feedback to Explore the
  Relationship Between Artists and Procedural Content Generation in Computer
  Game Development Tools}}.
\newblock \bibinfo{thesistype}{Ph.\,D. Dissertation}.
\newblock


\bibitem[Earle et~al\mbox{.}(2021)]%
        {earle2021learning}
\bibfield{author}{\bibinfo{person}{Sam Earle}, \bibinfo{person}{Maria Edwards},
  \bibinfo{person}{Ahmed Khalifa}, \bibinfo{person}{Philip Bontrager}, {and}
  \bibinfo{person}{Julian Togelius}.} \bibinfo{year}{2021}\natexlab{}.
\newblock \showarticletitle{Learning controllable content generators}. In
  \bibinfo{booktitle}{\emph{2021 IEEE Conference on Games (CoG)}}. IEEE,
  \bibinfo{pages}{1--9}.
\newblock


\bibitem[Earle et~al\mbox{.}(2022)]%
        {earle2022illuminating}
\bibfield{author}{\bibinfo{person}{Sam Earle}, \bibinfo{person}{Justin Snider},
  \bibinfo{person}{Matthew~C Fontaine}, \bibinfo{person}{Stefanos Nikolaidis},
  {and} \bibinfo{person}{Julian Togelius}.} \bibinfo{year}{2022}\natexlab{}.
\newblock \showarticletitle{Illuminating diverse neural cellular automata for
  level generation}. In \bibinfo{booktitle}{\emph{Proceedings of the Genetic
  and Evolutionary Computation Conference}}. \bibinfo{pages}{68--76}.
\newblock


\bibitem[Fan et~al\mbox{.}(2022)]%
        {fan2022minedojo}
\bibfield{author}{\bibinfo{person}{Linxi Fan}, \bibinfo{person}{Guanzhi Wang},
  \bibinfo{person}{Yunfan Jiang}, \bibinfo{person}{Ajay Mandlekar},
  \bibinfo{person}{Yuncong Yang}, \bibinfo{person}{Haoyi Zhu},
  \bibinfo{person}{Andrew Tang}, \bibinfo{person}{De-An Huang},
  \bibinfo{person}{Yuke Zhu}, {and} \bibinfo{person}{Anima Anandkumar}.}
  \bibinfo{year}{2022}\natexlab{}.
\newblock \showarticletitle{Minedojo: Building open-ended embodied agents with
  internet-scale knowledge}.
\newblock \bibinfo{journal}{\emph{arXiv preprint arXiv:2206.08853}}
  (\bibinfo{year}{2022}).
\newblock


\bibitem[Gao et~al\mbox{.}(2022)]%
        {gao2022procedural}
\bibfield{author}{\bibinfo{person}{Tianhan Gao}, \bibinfo{person}{Jin Zhang},
  {and} \bibinfo{person}{Qingwei Mi}.} \bibinfo{year}{2022}\natexlab{}.
\newblock \showarticletitle{Procedural Generation of Game Levels and Maps: A
  Review}. In \bibinfo{booktitle}{\emph{2022 International Conference on
  Artificial Intelligence in Information and Communication (ICAIIC)}}. IEEE,
  \bibinfo{pages}{050--055}.
\newblock


\bibitem[Gissl{\'e}n et~al\mbox{.}(2021)]%
        {gisslen2021adversarial}
\bibfield{author}{\bibinfo{person}{Linus Gissl{\'e}n}, \bibinfo{person}{Andy
  Eakins}, \bibinfo{person}{Camilo Gordillo}, \bibinfo{person}{Joakim
  Bergdahl}, {and} \bibinfo{person}{Konrad Tollmar}.}
  \bibinfo{year}{2021}\natexlab{}.
\newblock \showarticletitle{Adversarial reinforcement learning for procedural
  content generation}. In \bibinfo{booktitle}{\emph{2021 IEEE Conference on
  Games (CoG)}}. IEEE, \bibinfo{pages}{1--8}.
\newblock


\bibitem[Goodfellow et~al\mbox{.}(2020)]%
        {goodfellow2020generative}
\bibfield{author}{\bibinfo{person}{Ian Goodfellow}, \bibinfo{person}{Jean
  Pouget-Abadie}, \bibinfo{person}{Mehdi Mirza}, \bibinfo{person}{Bing Xu},
  \bibinfo{person}{David Warde-Farley}, \bibinfo{person}{Sherjil Ozair},
  \bibinfo{person}{Aaron Courville}, {and} \bibinfo{person}{Yoshua Bengio}.}
  \bibinfo{year}{2020}\natexlab{}.
\newblock \showarticletitle{Generative adversarial networks}.
\newblock \bibinfo{journal}{\emph{Commun. ACM}} \bibinfo{volume}{63},
  \bibinfo{number}{11} (\bibinfo{year}{2020}), \bibinfo{pages}{139--144}.
\newblock


\bibitem[Gravina et~al\mbox{.}(2019)]%
        {gravina2019procedural}
\bibfield{author}{\bibinfo{person}{Daniele Gravina}, \bibinfo{person}{Ahmed
  Khalifa}, \bibinfo{person}{Antonios Liapis}, \bibinfo{person}{Julian
  Togelius}, {and} \bibinfo{person}{Georgios~N Yannakakis}.}
  \bibinfo{year}{2019}\natexlab{}.
\newblock \showarticletitle{Procedural content generation through quality
  diversity}. In \bibinfo{booktitle}{\emph{2019 IEEE Conference on Games
  (CoG)}}. IEEE, \bibinfo{pages}{1--8}.
\newblock


\bibitem[Grbic et~al\mbox{.}(2021)]%
        {grbic2021evocraft}
\bibfield{author}{\bibinfo{person}{Djordje Grbic}, \bibinfo{person}{Rasmus~Berg
  Palm}, \bibinfo{person}{Elias Najarro}, \bibinfo{person}{Claire Glanois},
  {and} \bibinfo{person}{Sebastian Risi}.} \bibinfo{year}{2021}\natexlab{}.
\newblock \showarticletitle{Evocraft: A new challenge for open-endedness}. In
  \bibinfo{booktitle}{\emph{Applications of Evolutionary Computation: 24th
  International Conference, EvoApplications 2021, Held as Part of EvoStar 2021,
  Virtual Event, April 7--9, 2021, Proceedings 24}}. Springer,
  \bibinfo{pages}{325--340}.
\newblock


\bibitem[Green et~al\mbox{.}(2020)]%
        {green2020mario}
\bibfield{author}{\bibinfo{person}{Michael~Cerny Green},
  \bibinfo{person}{Luvneesh Mugrai}, \bibinfo{person}{Ahmed Khalifa}, {and}
  \bibinfo{person}{Julian Togelius}.} \bibinfo{year}{2020}\natexlab{}.
\newblock \showarticletitle{Mario level generation from mechanics using scene
  stitching}. In \bibinfo{booktitle}{\emph{2020 IEEE Conference on Games
  (CoG)}}. IEEE, \bibinfo{pages}{49--56}.
\newblock


\bibitem[Gumin(2016)]%
        {Gumin_Wave_Function_Collapse_2016}
\bibfield{author}{\bibinfo{person}{Maxim Gumin}.}
  \bibinfo{year}{2016}\natexlab{}.
\newblock \bibinfo{title}{{Wave Function Collapse Algorithm}}.
\newblock
\newblock
\urldef\tempurl%
\url{https://github.com/mxgmn/WaveFunctionCollapse}
\showURL{%
\tempurl}


\bibitem[Guo et~al\mbox{.}(2022)]%
        {guo2022neural}
\bibfield{author}{\bibinfo{person}{Xiang Guo}, \bibinfo{person}{Guanying Chen},
  \bibinfo{person}{Yuchao Dai}, \bibinfo{person}{Xiaoqing Ye},
  \bibinfo{person}{Jiadai Sun}, \bibinfo{person}{Xiao Tan}, {and}
  \bibinfo{person}{Errui Ding}.} \bibinfo{year}{2022}\natexlab{}.
\newblock \showarticletitle{Neural Deformable Voxel Grid for Fast Optimization
  of Dynamic View Synthesis}. In \bibinfo{booktitle}{\emph{Proceedings of the
  Asian Conference on Computer Vision}}. \bibinfo{pages}{3757--3775}.
\newblock


\bibitem[Guss et~al\mbox{.}(2021)]%
        {guss2021minerl}
\bibfield{author}{\bibinfo{person}{William~H Guss},
  \bibinfo{person}{Mario~Ynocente Castro}, \bibinfo{person}{Sam Devlin},
  \bibinfo{person}{Brandon Houghton}, \bibinfo{person}{Noboru~Sean Kuno},
  \bibinfo{person}{Crissman Loomis}, \bibinfo{person}{Stephanie Milani},
  \bibinfo{person}{Sharada Mohanty}, \bibinfo{person}{Keisuke Nakata},
  \bibinfo{person}{Ruslan Salakhutdinov}, {et~al\mbox{.}}}
  \bibinfo{year}{2021}\natexlab{}.
\newblock \showarticletitle{The minerl 2020 competition on sample efficient
  reinforcement learning using human priors}.
\newblock \bibinfo{journal}{\emph{arXiv preprint arXiv:2101.11071}}
  (\bibinfo{year}{2021}).
\newblock


\bibitem[Guzdial et~al\mbox{.}(2017)]%
        {guzdial2017visual}
\bibfield{author}{\bibinfo{person}{Matthew Guzdial}, \bibinfo{person}{Duri
  Long}, \bibinfo{person}{Christopher Cassion}, {and} \bibinfo{person}{Abhishek
  Das}.} \bibinfo{year}{2017}\natexlab{}.
\newblock \showarticletitle{Visual procedural content generation with an
  artificial abstract artist}. In \bibinfo{booktitle}{\emph{Proceedings of ICCC
  computational creativity and games workshop}}.
\newblock


\bibitem[Guzdial et~al\mbox{.}(2022a)]%
        {guzdial2022constraint}
\bibfield{author}{\bibinfo{person}{Matthew Guzdial}, \bibinfo{person}{Sam
  Snodgrass}, {and} \bibinfo{person}{Adam~J Summerville}.}
  \bibinfo{year}{2022}\natexlab{a}.
\newblock \showarticletitle{Constraint-Based PCGML Approaches}.
\newblock In \bibinfo{booktitle}{\emph{Procedural Content Generation via
  Machine Learning: An Overview}}. \bibinfo{publisher}{Springer},
  \bibinfo{pages}{51--66}.
\newblock


\bibitem[Guzdial et~al\mbox{.}(2022b)]%
        {guzdial2022pcgml}
\bibfield{author}{\bibinfo{person}{Matthew Guzdial}, \bibinfo{person}{Sam
  Snodgrass}, {and} \bibinfo{person}{Adam~J Summerville}.}
  \bibinfo{year}{2022}\natexlab{b}.
\newblock \showarticletitle{PCGML Process Overview}.
\newblock In \bibinfo{booktitle}{\emph{Procedural Content Generation via
  Machine Learning: An Overview}}. \bibinfo{publisher}{Springer},
  \bibinfo{pages}{35--49}.
\newblock


\bibitem[Guzdial et~al\mbox{.}(2022c)]%
        {guzdial2022procedural}
\bibfield{author}{\bibinfo{person}{Matthew Guzdial}, \bibinfo{person}{Sam
  Snodgrass}, {and} \bibinfo{person}{Adam~J Summerville}.}
  \bibinfo{year}{2022}\natexlab{c}.
\newblock \bibinfo{booktitle}{\emph{Procedural Content Generation Via Machine
  Learning: An Overview}}.
\newblock \bibinfo{publisher}{Springer}.
\newblock


\bibitem[Hao et~al\mbox{.}(2021)]%
        {hao2021gancraft}
\bibfield{author}{\bibinfo{person}{Zekun Hao}, \bibinfo{person}{Arun Mallya},
  \bibinfo{person}{Serge Belongie}, {and} \bibinfo{person}{Ming-Yu Liu}.}
  \bibinfo{year}{2021}\natexlab{}.
\newblock \showarticletitle{Gancraft: Unsupervised 3d neural rendering of
  minecraft worlds}. In \bibinfo{booktitle}{\emph{Proceedings of the IEEE/CVF
  International Conference on Computer Vision}}. \bibinfo{pages}{14072--14082}.
\newblock


\bibitem[Hendrikx et~al\mbox{.}(2013)]%
        {hendrikx2013procedural}
\bibfield{author}{\bibinfo{person}{Mark Hendrikx}, \bibinfo{person}{Sebastiaan
  Meijer}, \bibinfo{person}{Joeri Van Der~Velden}, {and}
  \bibinfo{person}{Alexandru Iosup}.} \bibinfo{year}{2013}\natexlab{}.
\newblock \showarticletitle{Procedural content generation for games: A survey}.
\newblock \bibinfo{journal}{\emph{ACM Transactions on Multimedia Computing,
  Communications, and Applications (TOMM)}} \bibinfo{volume}{9},
  \bibinfo{number}{1} (\bibinfo{year}{2013}), \bibinfo{pages}{1--22}.
\newblock


\bibitem[Jang et~al\mbox{.}(2017)]%
        {jang2017categorical}
\bibfield{author}{\bibinfo{person}{Eric Jang}, \bibinfo{person}{Shixiang Gu},
  {and} \bibinfo{person}{Ben Poole}.} \bibinfo{year}{2017}\natexlab{}.
\newblock \showarticletitle{Categorical Reparametrization with Gumble-Softmax}.
  In \bibinfo{booktitle}{\emph{International Conference on Learning
  Representations (ICLR 2017)}}. OpenReview. net.
\newblock


\bibitem[Jiang et~al\mbox{.}(2021)]%
        {jiang2021prioritized}
\bibfield{author}{\bibinfo{person}{Minqi Jiang}, \bibinfo{person}{Edward
  Grefenstette}, {and} \bibinfo{person}{Tim Rockt{\"a}schel}.}
  \bibinfo{year}{2021}\natexlab{}.
\newblock \showarticletitle{Prioritized level replay}. In
  \bibinfo{booktitle}{\emph{International Conference on Machine Learning}}.
  PMLR, \bibinfo{pages}{4940--4950}.
\newblock


\bibitem[Jiang et~al\mbox{.}(2022)]%
        {jiang2022learning}
\bibfield{author}{\bibinfo{person}{Zehua Jiang}, \bibinfo{person}{Sam Earle},
  \bibinfo{person}{Michael Green}, {and} \bibinfo{person}{Julian Togelius}.}
  \bibinfo{year}{2022}\natexlab{}.
\newblock \showarticletitle{Learning Controllable 3D Level Generators}. In
  \bibinfo{booktitle}{\emph{Proceedings of the 17th International Conference on
  the Foundations of Digital Games}}. \bibinfo{pages}{1--9}.
\newblock


\bibitem[Johnson et~al\mbox{.}(2016)]%
        {johnson2016malmo}
\bibfield{author}{\bibinfo{person}{Matthew Johnson}, \bibinfo{person}{Katja
  Hofmann}, \bibinfo{person}{Tim Hutton}, {and} \bibinfo{person}{David
  Bignell}.} \bibinfo{year}{2016}\natexlab{}.
\newblock \showarticletitle{The Malmo Platform for Artificial Intelligence
  Experimentation.}. In \bibinfo{booktitle}{\emph{Ijcai}}.
  \bibinfo{pages}{4246--4247}.
\newblock


\bibitem[Juliani et~al\mbox{.}(2019)]%
        {juliani2019obstacle}
\bibfield{author}{\bibinfo{person}{Arthur Juliani}, \bibinfo{person}{Ahmed
  Khalifa}, \bibinfo{person}{Vincent-Pierre Berges}, \bibinfo{person}{Jonathan
  Harper}, \bibinfo{person}{Ervin Teng}, \bibinfo{person}{Hunter Henry},
  \bibinfo{person}{Adam Crespi}, \bibinfo{person}{Julian Togelius}, {and}
  \bibinfo{person}{Danny Lange}.} \bibinfo{year}{2019}\natexlab{}.
\newblock \showarticletitle{Obstacle tower: A generalization challenge in
  vision, control, and planning}.
\newblock \bibinfo{journal}{\emph{arXiv preprint arXiv:1902.01378}}
  (\bibinfo{year}{2019}).
\newblock


\bibitem[Justesen et~al\mbox{.}(2018)]%
        {justesen2018illuminating}
\bibfield{author}{\bibinfo{person}{Niels Justesen},
  \bibinfo{person}{Ruben~Rodriguez Torrado}, \bibinfo{person}{Philip
  Bontrager}, \bibinfo{person}{Ahmed Khalifa}, \bibinfo{person}{Julian
  Togelius}, {and} \bibinfo{person}{Sebastian Risi}.}
  \bibinfo{year}{2018}\natexlab{}.
\newblock \showarticletitle{Illuminating generalization in deep reinforcement
  learning through procedural level generation}. In
  \bibinfo{booktitle}{\emph{NeurIPS Workshop on Deep Reinforcement Learning}}.
\newblock


\bibitem[Kanervisto et~al\mbox{.}(2022a)]%
        {kanervisto2022minerl}
\bibfield{author}{\bibinfo{person}{Anssi Kanervisto},
  \bibinfo{person}{Stephanie Milani}, \bibinfo{person}{Karolis Ramanauskas},
  \bibinfo{person}{Nicholay Topin}, \bibinfo{person}{Zichuan Lin},
  \bibinfo{person}{Junyou Li}, \bibinfo{person}{Jianing Shi},
  \bibinfo{person}{Deheng Ye}, \bibinfo{person}{Qiang Fu}, \bibinfo{person}{Wei
  Yang}, {et~al\mbox{.}}} \bibinfo{year}{2022}\natexlab{a}.
\newblock \showarticletitle{Minerl diamond 2021 competition: Overview, results,
  and lessons learned}.
\newblock \bibinfo{journal}{\emph{NeurIPS 2021 Competitions and Demonstrations
  Track}} (\bibinfo{year}{2022}), \bibinfo{pages}{13--28}.
\newblock


\bibitem[Kanervisto et~al\mbox{.}(2022b)]%
        {pmlr-v176-kanervisto22a}
\bibfield{author}{\bibinfo{person}{Anssi Kanervisto},
  \bibinfo{person}{Stephanie Milani}, \bibinfo{person}{Karolis Ramanauskas},
  \bibinfo{person}{Nicholay Topin}, \bibinfo{person}{Zichuan Lin},
  \bibinfo{person}{Junyou Li}, \bibinfo{person}{Jianing Shi},
  \bibinfo{person}{Deheng Ye}, \bibinfo{person}{Qiang Fu}, \bibinfo{person}{Wei
  Yang}, \bibinfo{person}{Weijun Hong}, \bibinfo{person}{Zhongyue Huang},
  \bibinfo{person}{Haicheng Chen}, \bibinfo{person}{Guangjun Zeng},
  \bibinfo{person}{Yue Lin}, \bibinfo{person}{Vincent Micheli},
  \bibinfo{person}{Eloi Alonso}, \bibinfo{person}{Fran\c{c}ois Fleuret},
  \bibinfo{person}{Alexander Nikulin}, \bibinfo{person}{Yury Belousov},
  \bibinfo{person}{Oleg Svidchenko}, {and} \bibinfo{person}{Aleksei Shpilman}.}
  \bibinfo{year}{2022}\natexlab{b}.
\newblock \showarticletitle{MineRL Diamond 2021 Competition: Overview, Results,
  and Lessons Learned}. In \bibinfo{booktitle}{\emph{Proceedings of the NeurIPS
  2021 Competitions and Demonstrations Track}}
  \emph{(\bibinfo{series}{Proceedings of Machine Learning Research},
  Vol.~\bibinfo{volume}{176})}, \bibfield{editor}{\bibinfo{person}{Douwe
  Kiela}, \bibinfo{person}{Marco Ciccone}, {and} \bibinfo{person}{Barbara
  Caputo}} (Eds.). \bibinfo{publisher}{PMLR}, \bibinfo{pages}{13--28}.
\newblock
\urldef\tempurl%
\url{https://proceedings.mlr.press/v176/kanervisto22a.html}
\showURL{%
\tempurl}


\bibitem[Karth and Smith(2019)]%
        {karth2019addressing}
\bibfield{author}{\bibinfo{person}{Isaac Karth} {and} \bibinfo{person}{Adam~M
  Smith}.} \bibinfo{year}{2019}\natexlab{}.
\newblock \showarticletitle{Addressing the fundamental tension of PCGML with
  discriminative learning}. In \bibinfo{booktitle}{\emph{Proceedings of the
  14th International Conference on the Foundations of Digital Games}}.
  \bibinfo{pages}{1--9}.
\newblock


\bibitem[Khalifa et~al\mbox{.}(2020)]%
        {khalifa2020pcgrl}
\bibfield{author}{\bibinfo{person}{Ahmed Khalifa}, \bibinfo{person}{Philip
  Bontrager}, \bibinfo{person}{Sam Earle}, {and} \bibinfo{person}{Julian
  Togelius}.} \bibinfo{year}{2020}\natexlab{}.
\newblock \showarticletitle{Pcgrl: Procedural content generation via
  reinforcement learning}. In \bibinfo{booktitle}{\emph{Proceedings of the AAAI
  Conference on Artificial Intelligence and Interactive Digital
  Entertainment}}, Vol.~\bibinfo{volume}{16}. \bibinfo{pages}{95--101}.
\newblock


\bibitem[K{\"u}ttler et~al\mbox{.}(2020)]%
        {kuttler2020nethack}
\bibfield{author}{\bibinfo{person}{Heinrich K{\"u}ttler},
  \bibinfo{person}{Nantas Nardelli}, \bibinfo{person}{Alexander Miller},
  \bibinfo{person}{Roberta Raileanu}, \bibinfo{person}{Marco Selvatici},
  \bibinfo{person}{Edward Grefenstette}, {and} \bibinfo{person}{Tim
  Rockt{\"a}schel}.} \bibinfo{year}{2020}\natexlab{}.
\newblock \showarticletitle{The nethack learning environment}.
\newblock \bibinfo{journal}{\emph{Advances in Neural Information Processing
  Systems}}  \bibinfo{volume}{33} (\bibinfo{year}{2020}),
  \bibinfo{pages}{7671--7684}.
\newblock


\bibitem[Lee and Chang(2022)]%
        {lee2022understanding}
\bibfield{author}{\bibinfo{person}{Han-Hung Lee} {and} \bibinfo{person}{Angel~X
  Chang}.} \bibinfo{year}{2022}\natexlab{}.
\newblock \showarticletitle{Understanding pure clip guidance for voxel grid
  nerf models}.
\newblock \bibinfo{journal}{\emph{arXiv preprint arXiv:2209.15172}}
  (\bibinfo{year}{2022}).
\newblock


\bibitem[Lee et~al\mbox{.}(2020)]%
        {lee2020precomputing}
\bibfield{author}{\bibinfo{person}{Vivian Lee}, \bibinfo{person}{Nathan
  Partlan}, {and} \bibinfo{person}{Seth Cooper}.}
  \bibinfo{year}{2020}\natexlab{}.
\newblock \showarticletitle{Precomputing Player Movement in Platformers for
  Level Generation with Reachability Constraints.}. In
  \bibinfo{booktitle}{\emph{AIIDE Workshops}}.
\newblock


\bibitem[Liapis et~al\mbox{.}(2013)]%
        {liapis2013designer}
\bibfield{author}{\bibinfo{person}{Antonios Liapis}, \bibinfo{person}{Georgios
  Yannakakis}, {and} \bibinfo{person}{Julian Togelius}.}
  \bibinfo{year}{2013}\natexlab{}.
\newblock \showarticletitle{Designer modeling for personalized game content
  creation tools}. In \bibinfo{booktitle}{\emph{Proceedings of the AAAI
  Conference on Artificial Intelligence and Interactive Digital
  Entertainment}}, Vol.~\bibinfo{volume}{9}. \bibinfo{pages}{11--16}.
\newblock


\bibitem[Liapis et~al\mbox{.}(2012)]%
        {liapis2012adapting}
\bibfield{author}{\bibinfo{person}{Antonios Liapis},
  \bibinfo{person}{Georgios~N Yannakakis}, {and} \bibinfo{person}{Julian
  Togelius}.} \bibinfo{year}{2012}\natexlab{}.
\newblock \showarticletitle{Adapting models of visual aesthetics for
  personalized content creation}.
\newblock \bibinfo{journal}{\emph{IEEE Transactions on Computational
  Intelligence and AI in Games}} \bibinfo{volume}{4}, \bibinfo{number}{3}
  (\bibinfo{year}{2012}), \bibinfo{pages}{213--228}.
\newblock


\bibitem[Lin et~al\mbox{.}(2014)]%
        {lin2014microsoft}
\bibfield{author}{\bibinfo{person}{Tsung-Yi Lin}, \bibinfo{person}{Michael
  Maire}, \bibinfo{person}{Serge Belongie}, \bibinfo{person}{James Hays},
  \bibinfo{person}{Pietro Perona}, \bibinfo{person}{Deva Ramanan},
  \bibinfo{person}{Piotr Doll{\'a}r}, {and} \bibinfo{person}{C~Lawrence
  Zitnick}.} \bibinfo{year}{2014}\natexlab{}.
\newblock \showarticletitle{Microsoft coco: Common objects in context}. In
  \bibinfo{booktitle}{\emph{European conference on computer vision}}. Springer,
  \bibinfo{pages}{740--755}.
\newblock


\bibitem[Liu et~al\mbox{.}(2021)]%
        {liu2021deep}
\bibfield{author}{\bibinfo{person}{Jialin Liu}, \bibinfo{person}{Sam
  Snodgrass}, \bibinfo{person}{Ahmed Khalifa}, \bibinfo{person}{Sebastian
  Risi}, \bibinfo{person}{Georgios~N Yannakakis}, {and} \bibinfo{person}{Julian
  Togelius}.} \bibinfo{year}{2021}\natexlab{}.
\newblock \showarticletitle{Deep learning for procedural content generation}.
\newblock \bibinfo{journal}{\emph{Neural Computing and Applications}}
  \bibinfo{volume}{33}, \bibinfo{number}{1} (\bibinfo{year}{2021}),
  \bibinfo{pages}{19--37}.
\newblock


\bibitem[LLC({[n.\,d.]})]%
        {planet-minecraft}
\bibfield{author}{\bibinfo{person}{Cyprezz LLC}.}
  \bibinfo{year}{[n.\,d.]}\natexlab{}.
\newblock \bibinfo{title}{Planet Minecraft Community: Creative fansite for
  everything minecraft!}
\newblock
\newblock
\newblock
\shownote{https://www.planetminecraft.com/}.


\bibitem[L{\'o}pez et~al\mbox{.}(2020)]%
        {lopez2020deep}
\bibfield{author}{\bibinfo{person}{Christian~E L{\'o}pez},
  \bibinfo{person}{James Cunningham}, \bibinfo{person}{Omar Ashour}, {and}
  \bibinfo{person}{Conrad~S Tucker}.} \bibinfo{year}{2020}\natexlab{}.
\newblock \showarticletitle{Deep reinforcement learning for procedural content
  generation of 3d virtual environments}.
\newblock \bibinfo{journal}{\emph{Journal of Computing and Information Science
  in Engineering}} \bibinfo{volume}{20}, \bibinfo{number}{5}
  (\bibinfo{year}{2020}).
\newblock


\bibitem[Medina et~al\mbox{.}(2023)]%
        {medina2023evolving}
\bibfield{author}{\bibinfo{person}{Alejandro Medina}, \bibinfo{person}{Melanie
  Richey}, \bibinfo{person}{Mark Mueller}, {and} \bibinfo{person}{Jacob
  Schrum}.} \bibinfo{year}{2023}\natexlab{}.
\newblock \showarticletitle{Evolving Flying Machines in Minecraft Using Quality
  Diversity}.
\newblock \bibinfo{journal}{\emph{arXiv preprint arXiv:2302.00782}}
  (\bibinfo{year}{2023}).
\newblock


\bibitem[Merino et~al\mbox{.}(2023a)]%
        {merino2023interactive}
\bibfield{author}{\bibinfo{person}{Timothy Merino}, \bibinfo{person}{M
  Charity}, {and} \bibinfo{person}{Julian Togelius}.}
  \bibinfo{year}{2023}\natexlab{a}.
\newblock \showarticletitle{Interactive Latent Variable Evolution for the
  Generation of Minecraft Structures}. In \bibinfo{booktitle}{\emph{Proceedings
  of the 18th International Conference on the Foundations of Digital Games}}.
  \bibinfo{pages}{1--8}.
\newblock


\bibitem[Merino et~al\mbox{.}(2023b)]%
        {merino2023five}
\bibfield{author}{\bibinfo{person}{Timothy Merino}, \bibinfo{person}{Roman
  Negri}, \bibinfo{person}{Dipika Rajesh}, \bibinfo{person}{M Charity}, {and}
  \bibinfo{person}{Julian Togelius}.} \bibinfo{year}{2023}\natexlab{b}.
\newblock \showarticletitle{The Five-Dollar Model: Generating Game Maps and
  Sprites from Sentence Embeddings}. In \bibinfo{booktitle}{\emph{Proceedings
  of the AAAI Conference on Artificial Intelligence and Interactive Digital
  Entertainment}}, Vol.~\bibinfo{volume}{19}. \bibinfo{pages}{107--115}.
\newblock


\bibitem[Milani et~al\mbox{.}(2020)]%
        {milani2020retrospective}
\bibfield{author}{\bibinfo{person}{Stephanie Milani}, \bibinfo{person}{Nicholay
  Topin}, \bibinfo{person}{Brandon Houghton}, \bibinfo{person}{William~H Guss},
  \bibinfo{person}{Sharada~P Mohanty}, \bibinfo{person}{Keisuke Nakata},
  \bibinfo{person}{Oriol Vinyals}, {and} \bibinfo{person}{Noboru~Sean Kuno}.}
  \bibinfo{year}{2020}\natexlab{}.
\newblock \showarticletitle{Retrospective analysis of the 2019 MineRL
  competition on sample efficient reinforcement learning}. In
  \bibinfo{booktitle}{\emph{NeurIPS 2019 Competition and Demonstration Track}}.
  PMLR, \bibinfo{pages}{203--214}.
\newblock


\bibitem[Mildenhall et~al\mbox{.}(2020)]%
        {mildenhall2020nerf}
\bibfield{author}{\bibinfo{person}{B Mildenhall}, \bibinfo{person}{PP
  Srinivasan}, \bibinfo{person}{M Tancik}, \bibinfo{person}{JT Barron},
  \bibinfo{person}{R Ramamoorthi}, {and} \bibinfo{person}{R Ng}.}
  \bibinfo{year}{2020}\natexlab{}.
\newblock \showarticletitle{Nerf: Representing scenes as neural radiance fields
  for view synthesis}. In \bibinfo{booktitle}{\emph{European conference on
  computer vision}}.
\newblock


\bibitem[Mohammad~Khalid et~al\mbox{.}(2022)]%
        {mohammad2022clip}
\bibfield{author}{\bibinfo{person}{Nasir Mohammad~Khalid},
  \bibinfo{person}{Tianhao Xie}, \bibinfo{person}{Eugene Belilovsky}, {and}
  \bibinfo{person}{Tiberiu Popa}.} \bibinfo{year}{2022}\natexlab{}.
\newblock \showarticletitle{CLIP-Mesh: Generating textured meshes from text
  using pretrained image-text models}. In \bibinfo{booktitle}{\emph{SIGGRAPH
  Asia 2022 Conference Papers}}. \bibinfo{pages}{1--8}.
\newblock


\bibitem[Mott et~al\mbox{.}(2019)]%
        {mott2019controllable}
\bibfield{author}{\bibinfo{person}{Justin Mott}, \bibinfo{person}{Saujas
  Nandi}, {and} \bibinfo{person}{Luke Zeller}.}
  \bibinfo{year}{2019}\natexlab{}.
\newblock \showarticletitle{Controllable and coherent level generation: A
  two-pronged approach}. In \bibinfo{booktitle}{\emph{Experimental AI in games
  workshop}}.
\newblock


\bibitem[Nair(2020)]%
        {nair2020using}
\bibfield{author}{\bibinfo{person}{Rohit Nair}.}
  \bibinfo{year}{2020}\natexlab{}.
\newblock \bibinfo{booktitle}{\emph{Using Raymarched shaders as environments in
  3D video games}}.
\newblock \bibinfo{publisher}{Drexel University}.
\newblock


\bibitem[Nelson et~al\mbox{.}(2016)]%
        {nelson2016rules}
\bibfield{author}{\bibinfo{person}{Mark~J Nelson}, \bibinfo{person}{Julian
  Togelius}, \bibinfo{person}{Cameron Browne}, {and} \bibinfo{person}{Michael
  Cook}.} \bibinfo{year}{2016}\natexlab{}.
\newblock \showarticletitle{Rules and mechanics}.
\newblock \bibinfo{journal}{\emph{Procedural Content Generation in Games}}
  (\bibinfo{year}{2016}), \bibinfo{pages}{99--121}.
\newblock


\bibitem[Parker-Holder et~al\mbox{.}(2022)]%
        {parker2022evolving}
\bibfield{author}{\bibinfo{person}{Jack Parker-Holder}, \bibinfo{person}{Minqi
  Jiang}, \bibinfo{person}{Michael Dennis}, \bibinfo{person}{Mikayel
  Samvelyan}, \bibinfo{person}{Jakob Foerster}, \bibinfo{person}{Edward
  Grefenstette}, {and} \bibinfo{person}{Tim Rockt{\"a}schel}.}
  \bibinfo{year}{2022}\natexlab{}.
\newblock \showarticletitle{Evolving curricula with regret-based environment
  design}. In \bibinfo{booktitle}{\emph{International Conference on Machine
  Learning}}. PMLR, \bibinfo{pages}{17473--17498}.
\newblock


\bibitem[Poole et~al\mbox{.}(2022)]%
        {poole2022dreamfusion}
\bibfield{author}{\bibinfo{person}{Ben Poole}, \bibinfo{person}{Ajay Jain},
  \bibinfo{person}{Jonathan~T Barron}, {and} \bibinfo{person}{Ben Mildenhall}.}
  \bibinfo{year}{2022}\natexlab{}.
\newblock \showarticletitle{Dreamfusion: Text-to-3d using 2d diffusion}.
\newblock \bibinfo{journal}{\emph{arXiv preprint arXiv:2209.14988}}
  (\bibinfo{year}{2022}).
\newblock


\bibitem[Radford et~al\mbox{.}(2021)]%
        {radford2021learning}
\bibfield{author}{\bibinfo{person}{Alec Radford}, \bibinfo{person}{Jong~Wook
  Kim}, \bibinfo{person}{Chris Hallacy}, \bibinfo{person}{Aditya Ramesh},
  \bibinfo{person}{Gabriel Goh}, \bibinfo{person}{Sandhini Agarwal},
  \bibinfo{person}{Girish Sastry}, \bibinfo{person}{Amanda Askell},
  \bibinfo{person}{Pamela Mishkin}, \bibinfo{person}{Jack Clark},
  {et~al\mbox{.}}} \bibinfo{year}{2021}\natexlab{}.
\newblock \showarticletitle{Learning transferable visual models from natural
  language supervision}. In \bibinfo{booktitle}{\emph{International conference
  on machine learning}}. PMLR, \bibinfo{pages}{8748--8763}.
\newblock


\bibitem[Risi and Togelius(2020)]%
        {risi2020increasing}
\bibfield{author}{\bibinfo{person}{Sebastian Risi} {and}
  \bibinfo{person}{Julian Togelius}.} \bibinfo{year}{2020}\natexlab{}.
\newblock \showarticletitle{Increasing generality in machine learning through
  procedural content generation}.
\newblock \bibinfo{journal}{\emph{Nature Machine Intelligence}}
  \bibinfo{volume}{2}, \bibinfo{number}{8} (\bibinfo{year}{2020}),
  \bibinfo{pages}{428--436}.
\newblock


\bibitem[Rombach et~al\mbox{.}(2022)]%
        {Rombach_2022_CVPR}
\bibfield{author}{\bibinfo{person}{Robin Rombach}, \bibinfo{person}{Andreas
  Blattmann}, \bibinfo{person}{Dominik Lorenz}, \bibinfo{person}{Patrick
  Esser}, {and} \bibinfo{person}{Bj\"orn Ommer}.}
  \bibinfo{year}{2022}\natexlab{}.
\newblock \showarticletitle{High-Resolution Image Synthesis With Latent
  Diffusion Models}. In \bibinfo{booktitle}{\emph{Proceedings of the IEEE/CVF
  Conference on Computer Vision and Pattern Recognition (CVPR)}}.
  \bibinfo{pages}{10684--10695}.
\newblock


\bibitem[Salge et~al\mbox{.}(2022)]%
        {salge2022impressions}
\bibfield{author}{\bibinfo{person}{Christoph Salge}, \bibinfo{person}{Claus
  Aranha}, \bibinfo{person}{Adrian Brightmoore}, \bibinfo{person}{Sean Butler},
  \bibinfo{person}{Rodrigo De~Moura~Canaan}, \bibinfo{person}{Michael Cook},
  \bibinfo{person}{Michael Green}, \bibinfo{person}{Hagen Fischer},
  \bibinfo{person}{Christian Guckelsberger}, \bibinfo{person}{Jupiter Hadley},
  {et~al\mbox{.}}} \bibinfo{year}{2022}\natexlab{}.
\newblock \showarticletitle{Impressions of the GDMC AI Settlement Generation
  Challenge in Minecraft}. In \bibinfo{booktitle}{\emph{Proceedings of the 17th
  International Conference on the Foundations of Digital Games}}.
  \bibinfo{pages}{1--16}.
\newblock


\bibitem[Salge et~al\mbox{.}(2018)]%
        {salge2018generative}
\bibfield{author}{\bibinfo{person}{Christoph Salge},
  \bibinfo{person}{Michael~Cerny Green}, \bibinfo{person}{Rodgrigo Canaan},
  {and} \bibinfo{person}{Julian Togelius}.} \bibinfo{year}{2018}\natexlab{}.
\newblock \showarticletitle{Generative design in minecraft (gdmc) settlement
  generation competition}. In \bibinfo{booktitle}{\emph{Proceedings of the 13th
  International Conference on the Foundations of Digital Games}}.
  \bibinfo{pages}{1--10}.
\newblock


\bibitem[Samvelyan et~al\mbox{.}(2021)]%
        {samvelyan2021minihack}
\bibfield{author}{\bibinfo{person}{Mikayel Samvelyan}, \bibinfo{person}{Robert
  Kirk}, \bibinfo{person}{Vitaly Kurin}, \bibinfo{person}{Jack Parker-Holder},
  \bibinfo{person}{Minqi Jiang}, \bibinfo{person}{Eric Hambro},
  \bibinfo{person}{Fabio Petroni}, \bibinfo{person}{Heinrich K{\"u}ttler},
  \bibinfo{person}{Edward Grefenstette}, {and} \bibinfo{person}{Tim
  Rockt{\"a}schel}.} \bibinfo{year}{2021}\natexlab{}.
\newblock \showarticletitle{Minihack the planet: A sandbox for open-ended
  reinforcement learning research}.
\newblock \bibinfo{journal}{\emph{arXiv preprint arXiv:2109.13202}}
  (\bibinfo{year}{2021}).
\newblock


\bibitem[Sarkar and Cooper(2020)]%
        {sarkar2020sequential}
\bibfield{author}{\bibinfo{person}{Anurag Sarkar} {and} \bibinfo{person}{Seth
  Cooper}.} \bibinfo{year}{2020}\natexlab{}.
\newblock \showarticletitle{Sequential segment-based level generation and
  blending using variational autoencoders}. In
  \bibinfo{booktitle}{\emph{Proceedings of the 15th International Conference on
  the Foundations of Digital Games}}. \bibinfo{pages}{1--9}.
\newblock


\bibitem[Sarkar et~al\mbox{.}(2020)]%
        {sarkar2020conditional}
\bibfield{author}{\bibinfo{person}{Anurag Sarkar}, \bibinfo{person}{Zhihan
  Yang}, {and} \bibinfo{person}{Seth Cooper}.} \bibinfo{year}{2020}\natexlab{}.
\newblock \showarticletitle{Conditional level generation and game blending}.
\newblock \bibinfo{journal}{\emph{arXiv preprint arXiv:2010.07735}}
  (\bibinfo{year}{2020}).
\newblock


\bibitem[Shaker et~al\mbox{.}(2016a)]%
        {shaker2016procedural}
\bibfield{author}{\bibinfo{person}{Noor Shaker}, \bibinfo{person}{Julian
  Togelius}, {and} \bibinfo{person}{Mark~J Nelson}.}
  \bibinfo{year}{2016}\natexlab{a}.
\newblock \showarticletitle{Procedural content generation in games}.
\newblock  (\bibinfo{year}{2016}).
\newblock


\bibitem[Shaker et~al\mbox{.}(2016b)]%
        {shaker2016mixed}
\bibfield{author}{\bibinfo{person}{Noor Shaker}, \bibinfo{person}{Julian
  Togelius}, \bibinfo{person}{Mark~J Nelson}, \bibinfo{person}{Antonios
  Liapis}, \bibinfo{person}{Gillian Smith}, {and} \bibinfo{person}{Noor
  Shaker}.} \bibinfo{year}{2016}\natexlab{b}.
\newblock \showarticletitle{Mixed-initiative content creation}.
\newblock \bibinfo{journal}{\emph{Procedural content generation in games}}
  (\bibinfo{year}{2016}), \bibinfo{pages}{195--214}.
\newblock


\bibitem[Shaker et~al\mbox{.}(2010)]%
        {shaker2010towards}
\bibfield{author}{\bibinfo{person}{Noor Shaker}, \bibinfo{person}{Georgios
  Yannakakis}, {and} \bibinfo{person}{Julian Togelius}.}
  \bibinfo{year}{2010}\natexlab{}.
\newblock \showarticletitle{Towards automatic personalized content generation
  for platform games}. In \bibinfo{booktitle}{\emph{Proceedings of the AAAI
  Conference on Artificial Intelligence and Interactive Digital
  Entertainment}}, Vol.~\bibinfo{volume}{6}. \bibinfo{pages}{63--68}.
\newblock


\bibitem[Singer et~al\mbox{.}(2022)]%
        {singer2022make}
\bibfield{author}{\bibinfo{person}{Uriel Singer}, \bibinfo{person}{Adam
  Polyak}, \bibinfo{person}{Thomas Hayes}, \bibinfo{person}{Xi Yin},
  \bibinfo{person}{Jie An}, \bibinfo{person}{Songyang Zhang},
  \bibinfo{person}{Qiyuan Hu}, \bibinfo{person}{Harry Yang},
  \bibinfo{person}{Oron Ashual}, \bibinfo{person}{Oran Gafni}, {et~al\mbox{.}}}
  \bibinfo{year}{2022}\natexlab{}.
\newblock \showarticletitle{Make-a-video: Text-to-video generation without
  text-video data}.
\newblock \bibinfo{journal}{\emph{arXiv preprint arXiv:2209.14792}}
  (\bibinfo{year}{2022}).
\newblock


\bibitem[Siper et~al\mbox{.}(2022)]%
        {siper2022path}
\bibfield{author}{\bibinfo{person}{Matthew Siper}, \bibinfo{person}{Ahmed
  Khalifa}, {and} \bibinfo{person}{Julian Togelius}.}
  \bibinfo{year}{2022}\natexlab{}.
\newblock \showarticletitle{Path of Destruction: Learning an Iterative Level
  Generator Using a Small Dataset}.
\newblock \bibinfo{journal}{\emph{arXiv preprint arXiv:2202.10184}}
  (\bibinfo{year}{2022}).
\newblock


\bibitem[Skjeltorp(2022)]%
        {skjeltorp20223d}
\bibfield{author}{\bibinfo{person}{Ole~Edvin Skjeltorp}.}
  \bibinfo{year}{2022}\natexlab{}.
\newblock \emph{\bibinfo{title}{3D Neural Cellular Automata-Simulating
  morphogenesis: Shape, color and behavior of three-dimensional structures}}.
\newblock \bibinfo{thesistype}{Master's\ thesis}.
\newblock


\bibitem[Smith and Mateas(2011)]%
        {smith2011answer}
\bibfield{author}{\bibinfo{person}{Adam~M Smith} {and} \bibinfo{person}{Michael
  Mateas}.} \bibinfo{year}{2011}\natexlab{}.
\newblock \showarticletitle{Answer set programming for procedural content
  generation: A design space approach}.
\newblock \bibinfo{journal}{\emph{IEEE Transactions on Computational
  Intelligence and AI in Games}} \bibinfo{volume}{3}, \bibinfo{number}{3}
  (\bibinfo{year}{2011}), \bibinfo{pages}{187--200}.
\newblock


\bibitem[Sudhakaran et~al\mbox{.}(2023)]%
        {sudhakaran2023mariogpt}
\bibfield{author}{\bibinfo{person}{Shyam Sudhakaran}, \bibinfo{person}{Miguel
  González-Duque}, \bibinfo{person}{Claire Glanois}, \bibinfo{person}{Matthias
  Freiberger}, \bibinfo{person}{Elias Najarro}, {and}
  \bibinfo{person}{Sebastian Risi}.} \bibinfo{year}{2023}\natexlab{}.
\newblock \bibinfo{title}{MarioGPT: Open-Ended Text2Level Generation through
  Large Language Models}.
\newblock
\newblock
\showeprint[arxiv]{2302.05981}~[cs.AI]


\bibitem[Sudhakaran et~al\mbox{.}(2021)]%
        {sudhakaran2021growing}
\bibfield{author}{\bibinfo{person}{Shyam Sudhakaran}, \bibinfo{person}{Djordje
  Grbic}, \bibinfo{person}{Siyan Li}, \bibinfo{person}{Adam Katona},
  \bibinfo{person}{Elias Najarro}, \bibinfo{person}{Claire Glanois}, {and}
  \bibinfo{person}{Sebastian Risi}.} \bibinfo{year}{2021}\natexlab{}.
\newblock \showarticletitle{Growing 3d artefacts and functional machines with
  neural cellular automata}.
\newblock \bibinfo{journal}{\emph{arXiv preprint arXiv:2103.08737}}
  (\bibinfo{year}{2021}).
\newblock


\bibitem[Summerville and Mateas(2015)]%
        {summerville2015sampling}
\bibfield{author}{\bibinfo{person}{Adam Summerville} {and}
  \bibinfo{person}{Michael Mateas}.} \bibinfo{year}{2015}\natexlab{}.
\newblock \showarticletitle{Sampling hyrule: Multi-technique probabilistic
  level generation for action role playing games}. In
  \bibinfo{booktitle}{\emph{Proceedings of the AAAI Conference on Artificial
  Intelligence and Interactive Digital Entertainment}},
  Vol.~\bibinfo{volume}{11}. \bibinfo{pages}{63--67}.
\newblock


\bibitem[Summerville et~al\mbox{.}(2018)]%
        {summerville2018procedural}
\bibfield{author}{\bibinfo{person}{Adam Summerville}, \bibinfo{person}{Sam
  Snodgrass}, \bibinfo{person}{Matthew Guzdial}, \bibinfo{person}{Christoffer
  Holmg{\aa}rd}, \bibinfo{person}{Amy~K Hoover}, \bibinfo{person}{Aaron
  Isaksen}, \bibinfo{person}{Andy Nealen}, {and} \bibinfo{person}{Julian
  Togelius}.} \bibinfo{year}{2018}\natexlab{}.
\newblock \showarticletitle{Procedural content generation via machine learning
  (PCGML)}.
\newblock \bibinfo{journal}{\emph{IEEE Transactions on Games}}
  \bibinfo{volume}{10}, \bibinfo{number}{3} (\bibinfo{year}{2018}),
  \bibinfo{pages}{257--270}.
\newblock


\bibitem[Sun et~al\mbox{.}(2022)]%
        {sun2022direct}
\bibfield{author}{\bibinfo{person}{Cheng Sun}, \bibinfo{person}{Min Sun}, {and}
  \bibinfo{person}{Hwann-Tzong Chen}.} \bibinfo{year}{2022}\natexlab{}.
\newblock \showarticletitle{Direct voxel grid optimization: Super-fast
  convergence for radiance fields reconstruction}. In
  \bibinfo{booktitle}{\emph{Proceedings of the IEEE/CVF Conference on Computer
  Vision and Pattern Recognition}}. \bibinfo{pages}{5459--5469}.
\newblock


\bibitem[Team et~al\mbox{.}(2023)]%
        {team2023human}
\bibfield{author}{\bibinfo{person}{Adaptive~Agent Team}, \bibinfo{person}{Jakob
  Bauer}, \bibinfo{person}{Kate Baumli}, \bibinfo{person}{Satinder Baveja},
  \bibinfo{person}{Feryal Behbahani}, \bibinfo{person}{Avishkar Bhoopchand},
  \bibinfo{person}{Nathalie Bradley-Schmieg}, \bibinfo{person}{Michael Chang},
  \bibinfo{person}{Natalie Clay}, \bibinfo{person}{Adrian Collister},
  {et~al\mbox{.}}} \bibinfo{year}{2023}\natexlab{}.
\newblock \showarticletitle{Human-Timescale Adaptation in an Open-Ended Task
  Space}.
\newblock \bibinfo{journal}{\emph{arXiv preprint arXiv:2301.07608}}
  (\bibinfo{year}{2023}).
\newblock


\bibitem[Team et~al\mbox{.}(2021)]%
        {team2021open}
\bibfield{author}{\bibinfo{person}{Open Ended~Learning Team},
  \bibinfo{person}{Adam Stooke}, \bibinfo{person}{Anuj Mahajan},
  \bibinfo{person}{Catarina Barros}, \bibinfo{person}{Charlie Deck},
  \bibinfo{person}{Jakob Bauer}, \bibinfo{person}{Jakub Sygnowski},
  \bibinfo{person}{Maja Trebacz}, \bibinfo{person}{Max Jaderberg},
  \bibinfo{person}{Michael Mathieu}, {et~al\mbox{.}}}
  \bibinfo{year}{2021}\natexlab{}.
\newblock \showarticletitle{Open-ended learning leads to generally capable
  agents}.
\newblock \bibinfo{journal}{\emph{arXiv preprint arXiv:2107.12808}}
  (\bibinfo{year}{2021}).
\newblock


\bibitem[Todd et~al\mbox{.}(2023)]%
        {todd2023level}
\bibfield{author}{\bibinfo{person}{Graham Todd}, \bibinfo{person}{Sam Earle},
  \bibinfo{person}{Muhammad~Umair Nasir}, \bibinfo{person}{Michael~Cerny
  Green}, {and} \bibinfo{person}{Julian Togelius}.}
  \bibinfo{year}{2023}\natexlab{}.
\newblock \showarticletitle{Level Generation Through Large Language Models}. In
  \bibinfo{booktitle}{\emph{Proceedings of the 18th International Conference on
  the Foundations of Digital Games}}. \bibinfo{pages}{1--8}.
\newblock


\bibitem[Togelius et~al\mbox{.}(2011)]%
        {togelius2011search}
\bibfield{author}{\bibinfo{person}{Julian Togelius},
  \bibinfo{person}{Georgios~N Yannakakis}, \bibinfo{person}{Kenneth~O Stanley},
  {and} \bibinfo{person}{Cameron Browne}.} \bibinfo{year}{2011}\natexlab{}.
\newblock \showarticletitle{Search-based procedural content generation: A
  taxonomy and survey}.
\newblock \bibinfo{journal}{\emph{IEEE Transactions on Computational
  Intelligence and AI in Games}} \bibinfo{volume}{3}, \bibinfo{number}{3}
  (\bibinfo{year}{2011}), \bibinfo{pages}{172--186}.
\newblock


\bibitem[Torrado et~al\mbox{.}(2020)]%
        {torrado2020bootstrapping}
\bibfield{author}{\bibinfo{person}{Ruben~Rodriguez Torrado},
  \bibinfo{person}{Ahmed Khalifa}, \bibinfo{person}{Michael~Cerny Green},
  \bibinfo{person}{Niels Justesen}, \bibinfo{person}{Sebastian Risi}, {and}
  \bibinfo{person}{Julian Togelius}.} \bibinfo{year}{2020}\natexlab{}.
\newblock \showarticletitle{Bootstrapping conditional gans for video game level
  generation}. In \bibinfo{booktitle}{\emph{2020 IEEE Conference on Games
  (CoG)}}. IEEE, \bibinfo{pages}{41--48}.
\newblock


\bibitem[Wang et~al\mbox{.}(2022)]%
        {wang2022attention}
\bibfield{author}{\bibinfo{person}{Peihao Wang}, \bibinfo{person}{Xuxi Chen},
  \bibinfo{person}{Tianlong Chen}, \bibinfo{person}{Subhashini Venugopalan},
  \bibinfo{person}{Zhangyang Wang}, {et~al\mbox{.}}}
  \bibinfo{year}{2022}\natexlab{}.
\newblock \showarticletitle{Is Attention All NeRF Needs?}
\newblock \bibinfo{journal}{\emph{arXiv preprint arXiv:2207.13298}}
  (\bibinfo{year}{2022}).
\newblock


\bibitem[Wang et~al\mbox{.}(2023)]%
        {wang2023f}
\bibfield{author}{\bibinfo{person}{Peng Wang}, \bibinfo{person}{Yuan Liu},
  \bibinfo{person}{Zhaoxi Chen}, \bibinfo{person}{Lingjie Liu},
  \bibinfo{person}{Ziwei Liu}, \bibinfo{person}{Taku Komura},
  \bibinfo{person}{Christian Theobalt}, {and} \bibinfo{person}{Wenping Wang}.}
  \bibinfo{year}{2023}\natexlab{}.
\newblock \showarticletitle{F\textsuperscript{2}-NeRF: Fast Neural Radiance
  Field Training with Free Camera Trajectories}.
\newblock \bibinfo{journal}{\emph{arXiv preprint arXiv:2303.15951}}
  (\bibinfo{year}{2023}).
\newblock


\bibitem[Watkins(2016)]%
        {watkins2016procedural}
\bibfield{author}{\bibinfo{person}{Ryan Watkins}.}
  \bibinfo{year}{2016}\natexlab{}.
\newblock \bibinfo{booktitle}{\emph{Procedural content generation for unity
  game development}}.
\newblock \bibinfo{publisher}{Packt Publishing Ltd}.
\newblock


\bibitem[Yannakakis and Togelius(2011)]%
        {yannakakis2011experience}
\bibfield{author}{\bibinfo{person}{Georgios~N Yannakakis} {and}
  \bibinfo{person}{Julian Togelius}.} \bibinfo{year}{2011}\natexlab{}.
\newblock \showarticletitle{Experience-driven procedural content generation}.
\newblock \bibinfo{journal}{\emph{IEEE Transactions on Affective Computing}}
  \bibinfo{volume}{2}, \bibinfo{number}{3} (\bibinfo{year}{2011}),
  \bibinfo{pages}{147--161}.
\newblock


\bibitem[Yates(2021)]%
        {yates2021use}
\bibfield{author}{\bibinfo{person}{Cristopher Yates}.}
  \bibinfo{year}{2021}\natexlab{}.
\newblock \emph{\bibinfo{title}{The use of Poisson Disc Distribution and A*
  Pathfinding for Procedural Content Generation in Minecraft}}.
\newblock \bibinfo{thesistype}{Ph.\,D. Dissertation}. \bibinfo{school}{Ph. D.
  Dissertation. Memorial University}.
\newblock


\bibitem[Zhang et~al\mbox{.}(2020)]%
        {zhang2020video}
\bibfield{author}{\bibinfo{person}{Hejia Zhang}, \bibinfo{person}{Matthew
  Fontaine}, \bibinfo{person}{Amy Hoover}, \bibinfo{person}{Julian Togelius},
  \bibinfo{person}{Bistra Dilkina}, {and} \bibinfo{person}{Stefanos
  Nikolaidis}.} \bibinfo{year}{2020}\natexlab{}.
\newblock \showarticletitle{Video game level repair via mixed integer linear
  programming}. In \bibinfo{booktitle}{\emph{Proceedings of the AAAI Conference
  on Artificial Intelligence and Interactive Digital Entertainment}},
  Vol.~\bibinfo{volume}{16}. \bibinfo{pages}{151--158}.
\newblock


\bibitem[Zook and Riedl(2014)]%
        {zook2014generating}
\bibfield{author}{\bibinfo{person}{Alexander Zook} {and}
  \bibinfo{person}{Mark~O Riedl}.} \bibinfo{year}{2014}\natexlab{}.
\newblock \showarticletitle{Generating and adapting game mechanics}. In
  \bibinfo{booktitle}{\emph{Proceedings of the 2014 Foundations of Digital
  Games Workshop on Procedural Content Generation in Games}}.
\newblock


\end{thebibliography}

\appendix

\section{Minecraft Textures}

The set of Minecraft textures used to imitate in-game blocks within the neural rendering engine is displayed in Table~\ref{tab:textures}.

\begin{table*}[]
\centering
\begin{tabularx}{\linewidth}{c|X|c|X|c|X}
\toprule
 block & texture & block & texture & block & texture \\
\toprule
 log oak & \includegraphics{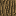}  \includegraphics{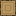}& 
 stone & \includegraphics{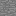}  &
 dirt & \includegraphics[]{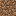}  \\
 brick & \includegraphics[]{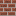} &
 clay & \includegraphics{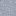} &
 snow & \includegraphics{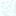} 
 \\
\makecell{glazed \\terracotta light blue }& \includegraphics{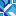} &  
\makecell{glazed \\terracotta yellow} & \includegraphics{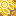} & 
\makecell{redstone\\block} & \includegraphics{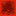} 
 \\
 gold block & \includegraphics{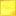} & 
 iron block & \includegraphics{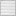} & 
 \makecell{diamond\\block} & \includegraphics{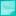} 
 \\
\makecell{emerald\\block} & \includegraphics{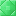}  &
 cobblestone & \includegraphics{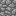} & 
 slime & \includegraphics{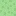}  \\
 \bottomrule
\end{tabularx}
\caption{In-game blocks and textures used by \ours{}}
\label{tab:textures}
\end{table*}





\section{\ours{} Generations}

\begin{figure*}[h!]
\centering
\begin{subfigure}[b]{\textwidth}
\includegraphics[width=.32\textwidth]{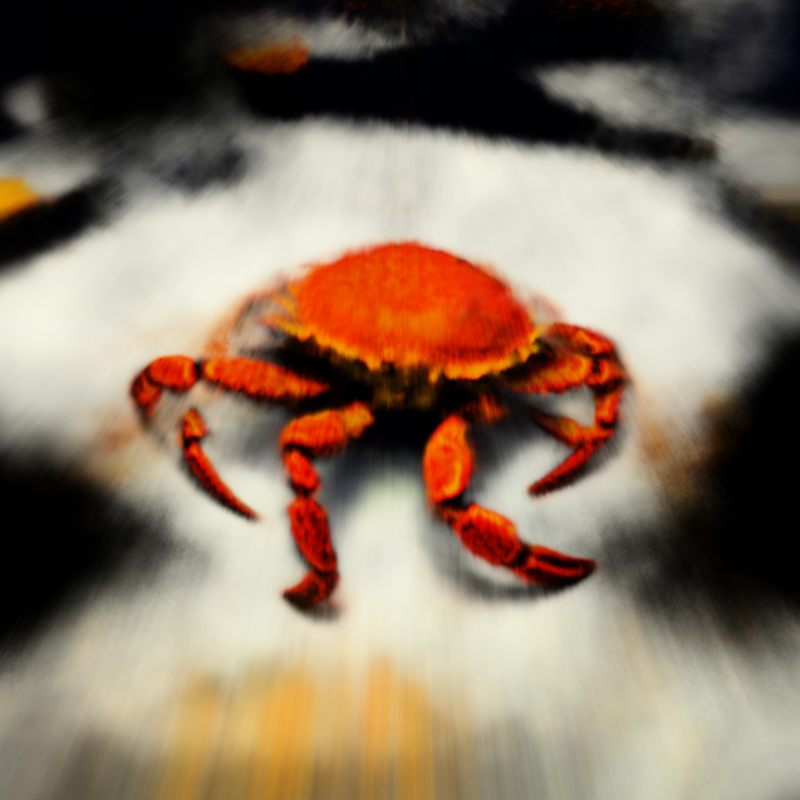}
\includegraphics[width=.32\textwidth]{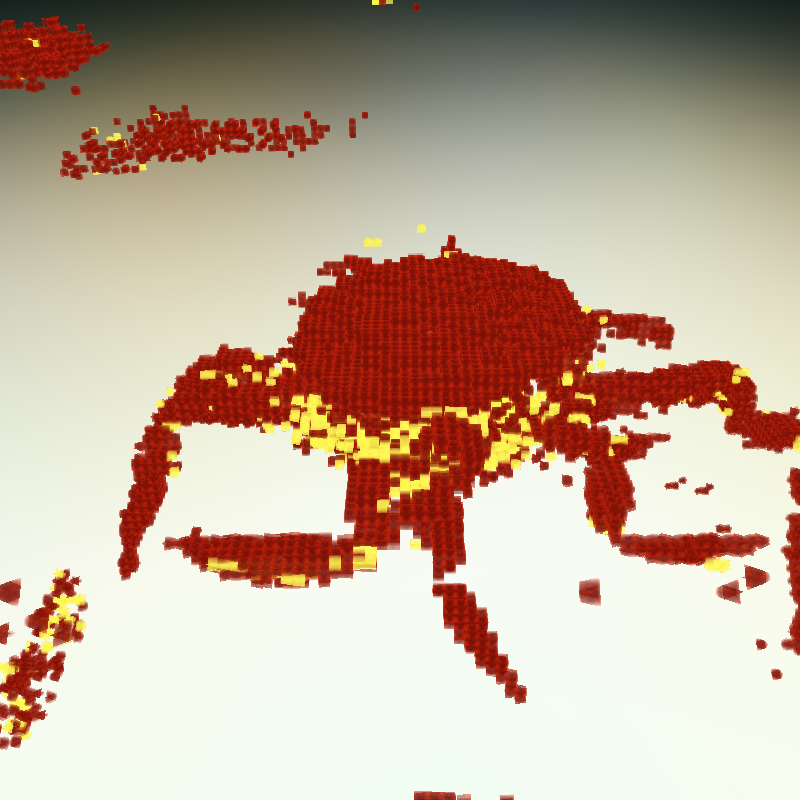}
\includegraphics[width=.32\textwidth]{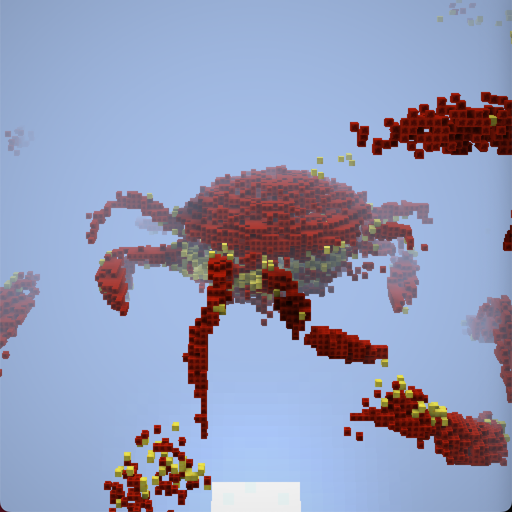}
\caption*{\textit{A pile of crab is seasoned and well cooked}}
\end{subfigure}\\
\begin{subfigure}[b]{\textwidth}
\begin{subfigure}[b]{.32\textwidth}
\includegraphics[width=\textwidth]{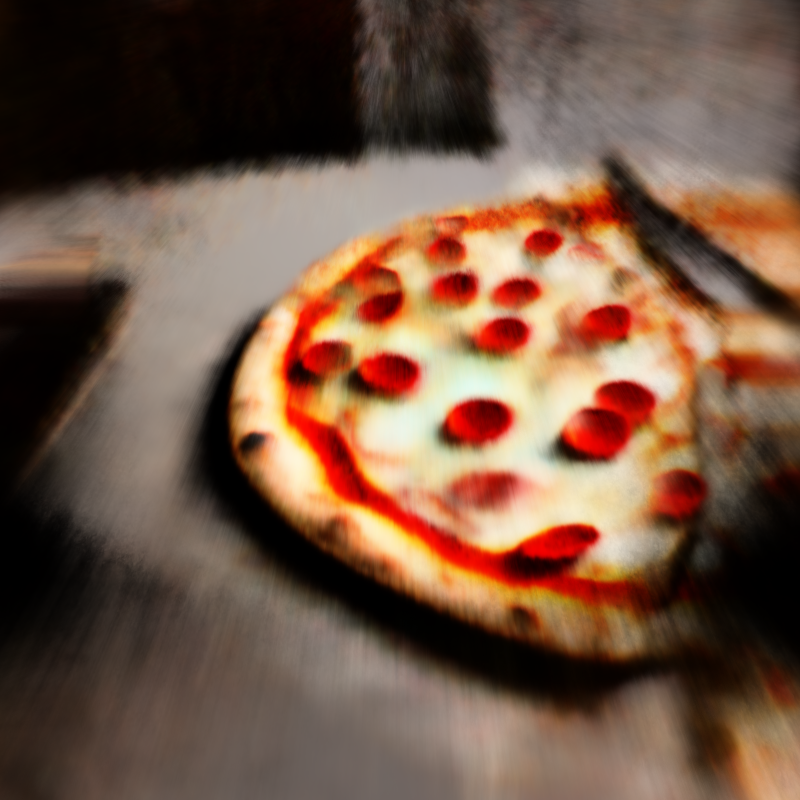}
\caption*{\unconstrained{}}
\end{subfigure}
\begin{subfigure}[b]{.32\textwidth}
\includegraphics[width=\textwidth]{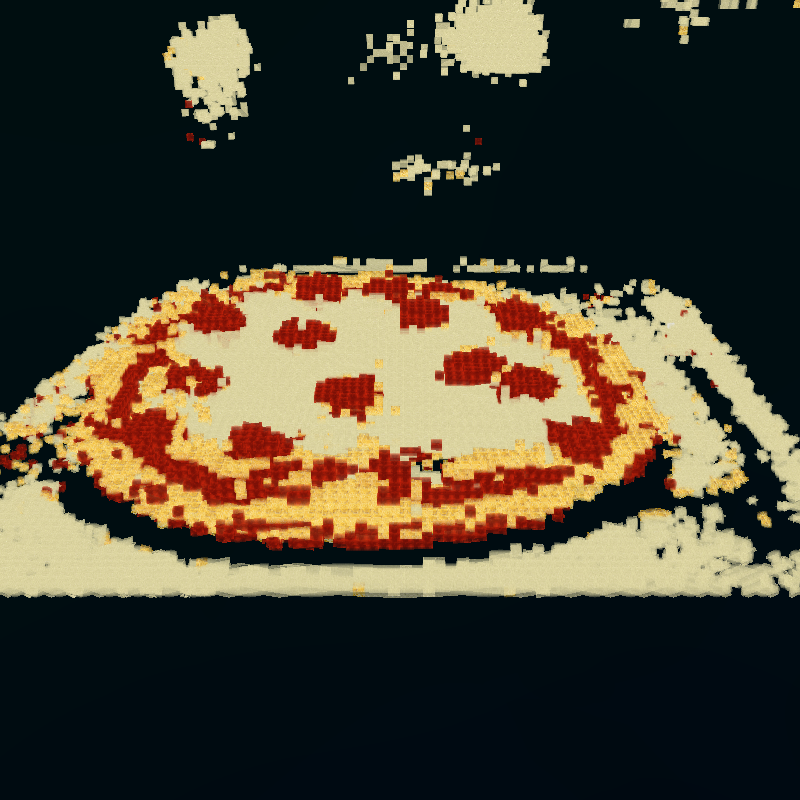}
\caption*{$N=100$}
\end{subfigure}
\begin{subfigure}[b]{.32\textwidth}
\includegraphics[width=\textwidth]{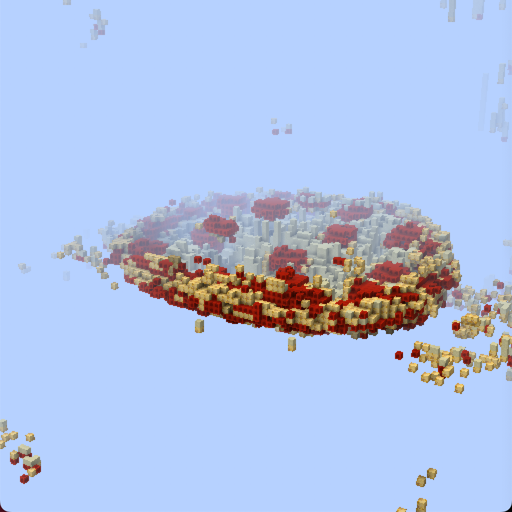}
\caption*{$N=100$, in-game}
\end{subfigure}
\caption*{\textit{A pizza and fork on a tray on the table}}
\end{subfigure}
\caption{\ours{} output given COCO dataset captions, viewed in-game (right), by neural rendering (middle), and standard text-guided NeRF output (left) given the same prompts.}
\label{fig:coco_unconstrained}
\end{figure*}

\begin{figure*}[h!]
    \begin{subfigure}[t]{\textwidth}
    \begin{subfigure}[t]{.33\textwidth}
        \includegraphics[width=\textwidth]{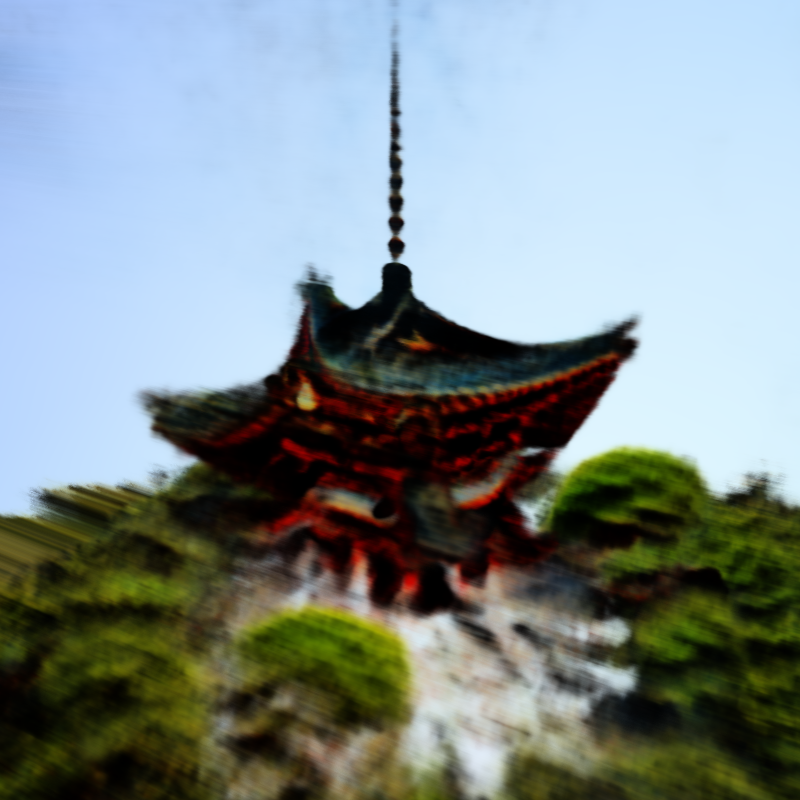}
    \end{subfigure}
    \begin{subfigure}[t]{.33\textwidth}
        \includegraphics[width=\textwidth]{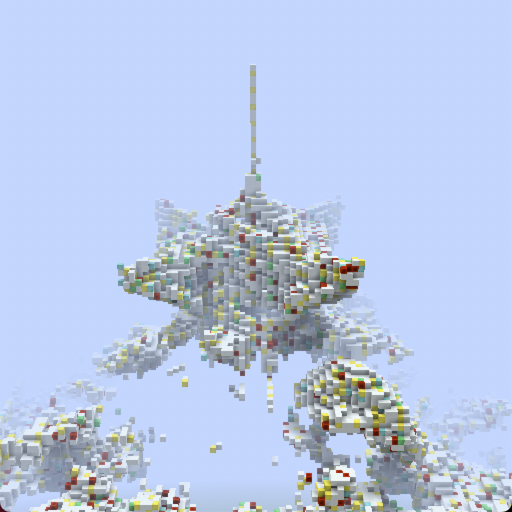}
    \end{subfigure}
    \begin{subfigure}[t]{.33\textwidth}
        \includegraphics[width=\textwidth]{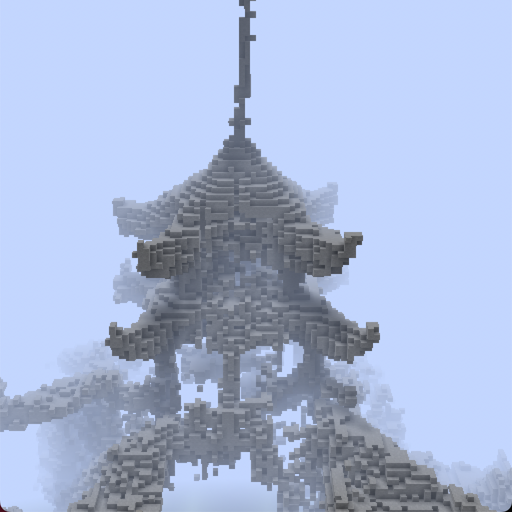}
    \end{subfigure}
    \caption{\textit{a japanese temple}}
        \begin{subfigure}[t]{.33\textwidth}
        \includegraphics[width=\textwidth]{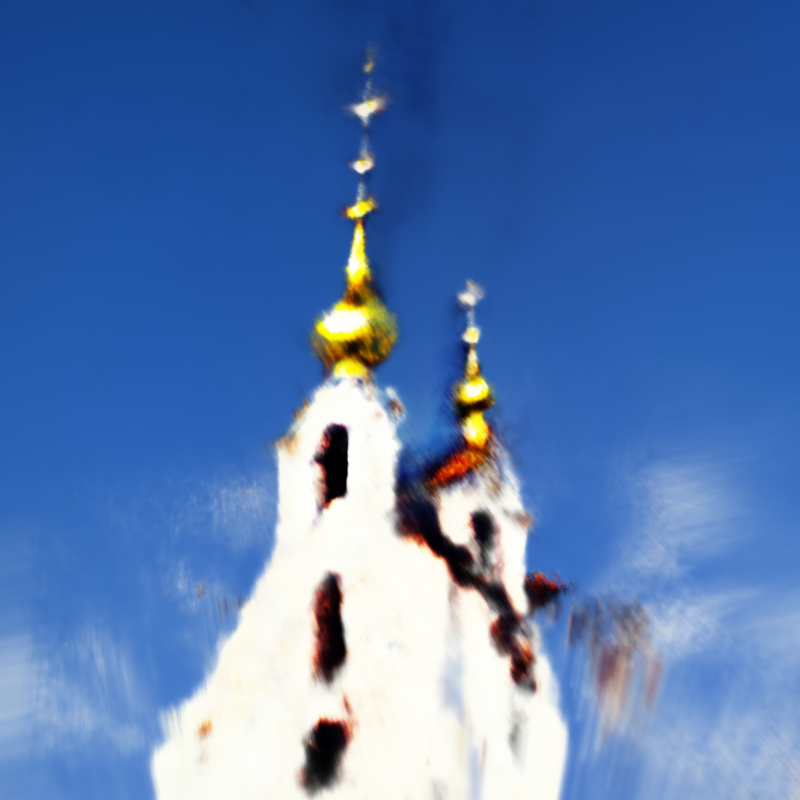}
    \caption*{Unconstrained text-guided NeRF}
    \end{subfigure}
    \begin{subfigure}[t]{.33\textwidth}
        \includegraphics[width=\textwidth]{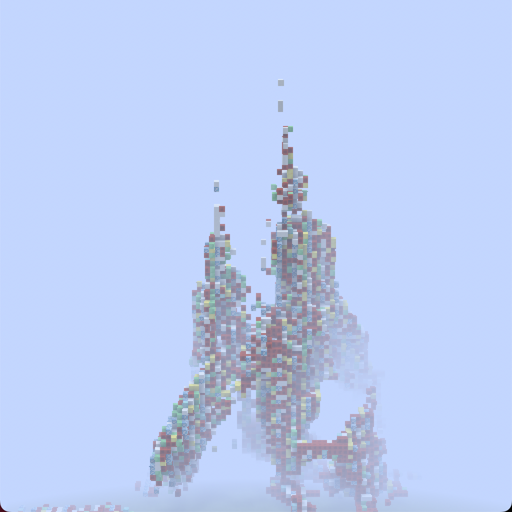}
    \caption*{text-guided NeRF with post hoc quantization using nearest neighbor mapping}
    \end{subfigure}
    \begin{subfigure}[t]{.33\textwidth}
        \includegraphics[width=\textwidth]{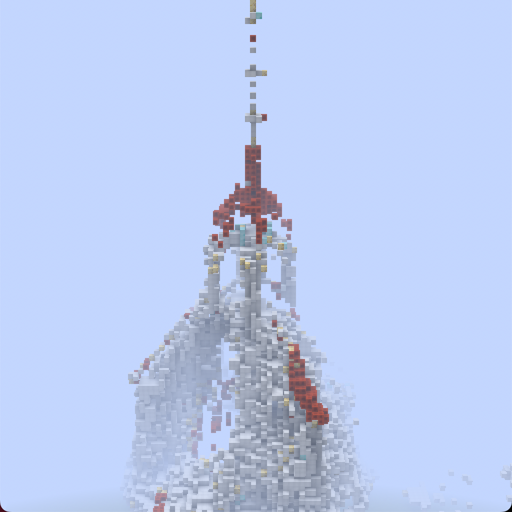}
    \caption*{\ours{}}
    \end{subfigure}
    \caption{\textit{church of the annunciation of the blessed virgin mary}}
    \end{subfigure}
\caption{\ours{} directly optimizes a representation using Minecraft blocks, leading to a more faithful reconstruction of the text prompt than results from post-processing the output of an \unconstrained{}, which leads to less coherent block type selection and structure topology. Captions from the planet minecraft dataset.}
\label{fig:postprocess_baseline}
\end{figure*}

\section{Limitations}

In some cases, the model uses negative space to represent an object, modulating the background texture to a particular color, then occluding parts of it with foreground blocks/density, to give it an apparent shape. This undesirable swapping of roles between foreground/background models may be more likely to occur in the quantized NeRF: whereas certain colors or textures may be difficult or impossible to replicate using the provided blocks, the background MLP remains unconstrained. To mitigate this, future work could investigate constraining the background MLP to only use 2D projections of ``distant'' game assets.

Another potential issue is the lack of semantic grounding with respect to block types. For example the model may just as well satisfy the prompt ``large medieval ship'' by using a combination of dirt and redstone, as with actual wooden logs or planks, so long as these give the \textit{appearance} of wood. Our preliminary work on functional constraints suggests that this particular problem can be addressed by setting per-block-type targets (e.g. requiring 0\% dirt and 50\% wood blocks), but a more general approach might lie in ``demonstrating'' what each block type should be used to represent by adding this information in the prompt.

Whereas traditional NeRFs can model lighting and shadows, this is not the case in \ours{}, where the color at each point in 3D space is derived directly from a voxel grid corresponding to the in-game appearance of a Minecraft block. When structures are rendered inside the neural engine, they thus appear ``flat'' in contrast to the kind of shadow and lighting effects that appear in the Minecraft game engine. Ideally, we could train an auxiliary model to mimic the effects of in-game lighting, for example by training it on paired datapoints of 3D block grids, and their appearance in-game at various angles. This learned renderer could replace the differentiable raycasting component of the NeRF pipeline (as in \cite{wang2022attention}), further sparing us from having to re-implement the rendering of irregular game objects such as plants and glass.

\ours{} is currently too slow to be feasibly used in an online player-environment generator loop, taking a few hours to generate a single structure. Future versions could benefit from recent and future speed improvements in NeRFs~\cite{wang2023f,guo2022neural,sun2022direct,zhang2020video}. Alternatively, it could be leveraged to generate a training set for a conditional, guidance-free generative models of game worlds.

\section{Planet Minecraft Dataset}
\label{sec:planet_mc_dataset}

To test \ours{}'s ability to generate environments specific to the domain for which it was designed, we source text prompts from Planet Minecraft, a fan-operated site where users can upload and share custom content. We consider a subset of assets uploaded to the ``Maps'' category in 2016 (the year in which the most such assets were uploaded), and select the top 150 maps of this subset as measued by the number of user downloads. The prompts correspond to the names of these assets. We do not collect the assets themselves or any further data from the site.


The set of prompts scraped from Planet Minecraft is given below:
\begin{figure*}[]
\texttt{
| paris   eiffel tower | la valle dor by mrbatou download cinematic | reims cathedral | summit creative house | mexican hacjenda | coruscant senate building | from my house to yours merry christmas | the ziggurat | polaris skyscraper 25 | the craftsmans abode   pmc solo contest 4 | rustic fantasy house timelapse download | skyscraper 31 ias | chateau de silveberg | fuminsh the city that never sleeps | ontario tower | dirt modern house | farin rocks | elven tower of the wise | distorsion chunk challenge | tours thiers   nancy   france skyscraper 3 ias | luxury beach house | space lighthouse | central place modern office complex | minecraft is a small world | tiger ii 101 scale | shurwyth snowlands   download weareconquest | bridges | jurassic world v2 for jurassicraft 20 minecraft dinosaurs jurassic park isla nublar | brynwalda survival map | icarly set and nickelodeon studio | battle of hoth map echo base star wars hoth map | dream | the little castle nebelburg | steampunk island | land of azorth | jaws ride and amity village | avatar   base | quartz tower 1 | small modern house 3 full interior | minecraft disneyland 1965 | 2012 skyrim | greenfield project   neoclassical house | 2012 sea dogs village | eternal haven | patronis | japanese temple | star labs the flash cw | hub spawn | church of the annunciation of the blessed virgin mary   inowrocaw | space needle accurate | the dark city   hokkaido | a watch tower inspired by the game firewatch | undertale | white snowy castle | avengers tower | fantme villa   modern house 2 | abandoned wild west | greenfield building   vista creek elementary school | fantasy bundle level 25 special | a modern house 1 | server spawn by infro\_ | fantasy inspired village danjgames | classic american farm | the builders shrine   chunk challange | der eisendrache | modern house by real architect | ahzvels hq download   re upload | download a medieval detached farm showcase | wg tower   vice city | calypso   a modern villa | tf2 egypt | minigames map   atlantis | large medieval ship | small cabin | sustainable city | large medieval ship | sequoia valley 30 | kraehenfels survival version | small hospital | panem mc 1st quarter quell arena download | a desperate and lonely wizards tower pmc chunk challenge entry lore | paper mariocolour splash port prisma | skyscraper   planus | hollywood residence | a nordic mountain village | old wizards tree mansion series 2 build 1read description | hidden in the sand | five nights at candys 2   roleplay map | castle ardor | epic server spawn | medium medieval home | medeival windmill 1102 | grand stadium pixelmon | prison mine 2 with download | babylon gardens | the new world trade center 11 | 30days   day 28   orcish butchers slaughter house | old west home | ark labs outpost 26 | icebornminigames map | grand university | medieval mountain castle | minecraft maps | victorian manor | 30days   day 8   dwarven entrance | spherical greenhouse 18 19 110 | huge minecraft server spawn airidale | big cottage | greenfield   typical victorian | old fortress | modern condoapartment | small castle | ss tropic a custom by prestogo | two story old west shop | citadel hill fort george | batman arkham asylum | forest cottage | mountain temple | the conjuring | northwich | the amazing word of gumball the wattersons house 18 | five nights at freddys minecraft map 19 and above | fnaf roleplay map | middle eastern farm | modern house build sorry about no music in the video | a medieval farm | woodland mansion | survival map   jairus isle | mario kartgba bowsers castle 2 | the scandinavian townhouse | mesa fort | three cool wood structures | brenttwood estates | toy shop 192 | dota 2 | server shop | store   fletchers retreat | custom realistic terrain | the piggy sphinx | shubbles castle building contest | mountain housecastle 1 | roman outpost | fallout 4 red rocket | fantasy house | savanna village in the sky cinematic download | kahuai city | minecraft lets build timelapse   fantasy   update 12   over hanging house | kent regional airport | medieval house | redstone bunker 13 redstone creations version 1
}
\caption*{\ours{} prompts from Planet Minecraft}
\end{figure*}

\end{document}